\documentclass[prd, a4paper, amsfonts, amssymb, amsmath, reprint, showkeys, nofootinbib, twoside]{revtex4-1}

\usepackage[english]{babel}
\usepackage[utf8]{inputenc}
\usepackage[toc,page]{appendix}
\usepackage{verbatim}
\usepackage{tikz}
\usepackage{placeins}
\usepackage[colorinlistoftodos,color=green!40,prependcaption]{todonotes}
\usepackage{amsthm}
\usepackage{mathtools}
\usepackage{physics}
\usepackage{xcolor}
\usepackage{graphicx}
\usepackage[left=23mm,right=13mm,top=35mm,columnsep=15pt]{geometry} 
\usepackage{adjustbox}
\usepackage{placeins}
\usepackage[T1]{fontenc}
\usepackage{lipsum}
\usepackage{csquotes}
\usepackage{subfig}
\usepackage{caption}
\usepackage{comment}
\usepackage{wrapfig}
\usepackage{pinlabel, color}
\usepackage{multirow}
\usepackage{array}

\usepackage[pdftex, pdftitle={Article}, pdfauthor={Author}]{hyperref} 
\usepackage{array}
\newcolumntype{P}[1]{>{\centering\arraybackslash}p{#1}}
\newcolumntype{M}[1]{>{\centering\arraybackslash}m{#1}}

\newcommand{\etal}{\emph{et al.} }

\begin{document}
\title{Holography for Ising spins on the hyperbolic plane}

\author{Muhammad Asaduzzaman}
\email{masaduzz@syr.edu}
\affiliation{Department of Physics, Syracuse University, Syracuse, NY 13244, USA.}

\author{Simon Catterall}
\email{smcatter@syr.edu}
\affiliation{Department of Physics, Syracuse University, Syracuse, NY 13244, USA.}

\author{Jay Hubisz}
\email{jhubisz@syr.edu}
\affiliation{Department of Physics, Syracuse University, Syracuse, NY 13244, USA.}

\author{Roice Nelson}
\email{roice3@gmail.com}
\affiliation{Department of Physics, Syracuse University, Syracuse, NY 13244, USA.}

\author{Judah Unmuth-Yockey}
\email{jfunmuthyockey@gmail.com}
\affiliation{Department of Theoretical Physics, Fermi National Accelerator Laboratory,
Batavia, IL, USA.}
\date{\today} 

\begin{abstract}
Motivated by the AdS/CFT correspondence, we use Monte Carlo simulation to investigate the Ising model formulated on tessellations of the two-dimensional hyperbolic disk. We focus in particular on the behavior of boundary-boundary correlators, which exhibit power-law scaling both below and above the bulk critical temperature indicating scale invariance of the boundary theory at any temperature. This conclusion is strengthened by a finite-size scaling analysis of the boundary susceptibility which yields a scaling exponent consistent with the scaling dimension extracted from the boundary correlation function. This observation provides evidence that the connection between continuum boundary conformal symmetry and isometries of the bulk hyperbolic space survives for simple interacting field theories even when the bulk is approximated by a 
discrete tessellation.
\end{abstract}

\maketitle

\section{Introduction} \label{sec1}
The AdS/CFT correspondence~\cite{Maldacena:1997re,Witten:1998qj,KLEBANOV199989} is an immensely powerful tool in the theorists and phenomenologists arsenal, providing a duality dictionary between strongly coupled $d$-dimensional critical systems and weakly coupled $d+1$ dimensional gravitational theories on a negatively curved background.  This duality is holographic in nature, with the $d$-dimensional non-gravitational conformal theory residing on the boundary of $AdS_{d+1}$.

Key to this duality is the relation between boundary and bulk distance.  In a space with negative curvature, in this discussion presumed to be rigid, the length of a bulk geodesic between two boundary points is logarithmic in the distance for a path  restricted to the boundary hypersurface.  Specifically, $\sqrt{-k}\, d_\text{bulk} \sim \log (\sqrt{-k}\, d_\text{boundary})$, where $k$ is the curvature of the hyperbolic space.

At generic points in coupling space, a bulk $d+1$ dimensional theory will be gapped at scale $\mu$, and correlation functions will decay exponentially, falling off like $e^{-\mu d_\text{bulk}}$.  If expressed in terms of distance along the boundary, that same correlator will instead fall off as a power-law: $\left(d_\text{boundary}\right)^{-\mu}$.  Hence, the geometry of the bulk thus dictates that conformal behavior on the boundary will be robust as physical parameters of the bulk theory are varied.

Surprisingly, the most remarkable feature of this duality---the robustness of critical behavior of the boundary theory---persists even when only crude features of the geometry and field content are maintained.  For example: in previous work~\cite{Asaduzzaman:2020hjl}, we studied a model of a massive free scalar field propagating on tessellations of two and three-dimensional hyperbolic space.  Despite strong lattice artifacts associated with finite lattice spacing and finite volume, the boundary lattice theory displays the usual features of conformality, exhibiting power-law fall-off of boundary-to-boundary correlators with boundary distance, where the inferred scaling dimensions match precisely with continuum analysis.

Here we take the story further, exploring a simple but strongly interacting lattice quantum field theory living on hyperbolic space. The Ising model on the discretized Poincar{\'e} disk exhibits phase structure much like the flat space 2D Ising model: it possesses gapped ferromagnetic and paramagnetic phases separated by a phase transition. However, we will show that boundary-to-boundary correlators exhibit signals of criticality for a wide range of temperatures, as predicted by properties of the geometry. 

This study is novel in that it explores the impact of strongly coupled bulk physics on the AdS/CFT correspondence.  The correspondence is more typically understood in the regime of large $N$ CFTs with weakly coupled AdS duals, so this paper probes the correspondence in a regime uniquely suited to the tools of lattice quantum field theory.

To form our conclusions, we have used Monte Carlo
simulation and measured both boundary and bulk observables---with emphasis on the boundary observables---for a range
of temperatures. We furthermore show that this behavior can be understood theoretically
using a combination of high-temperature
expansion and duality arguments.

The organization of the paper is described here for the reader. In section~\ref{sec_bulk}, we describe the model in detail and discuss the
bulk phase structure.  In section~\ref{boundary} we investigate the behavior of boundary observables---the two-point correlation function and thermodynamics. We summarise our main results and discuss the prospects of our work in section~\ref{sec:summary}. Additional bulk results are added in appendix~\ref{App1} and results for boundary observables in a dual model are added in appendix~\ref{App2}. 

\section{The model and bulk phase structure}{\label{sec_bulk}}

\begin{figure}[!t]
    \subfloat[$\{3,7\}$ disk]{%
      \includegraphics[width=0.8\linewidth]{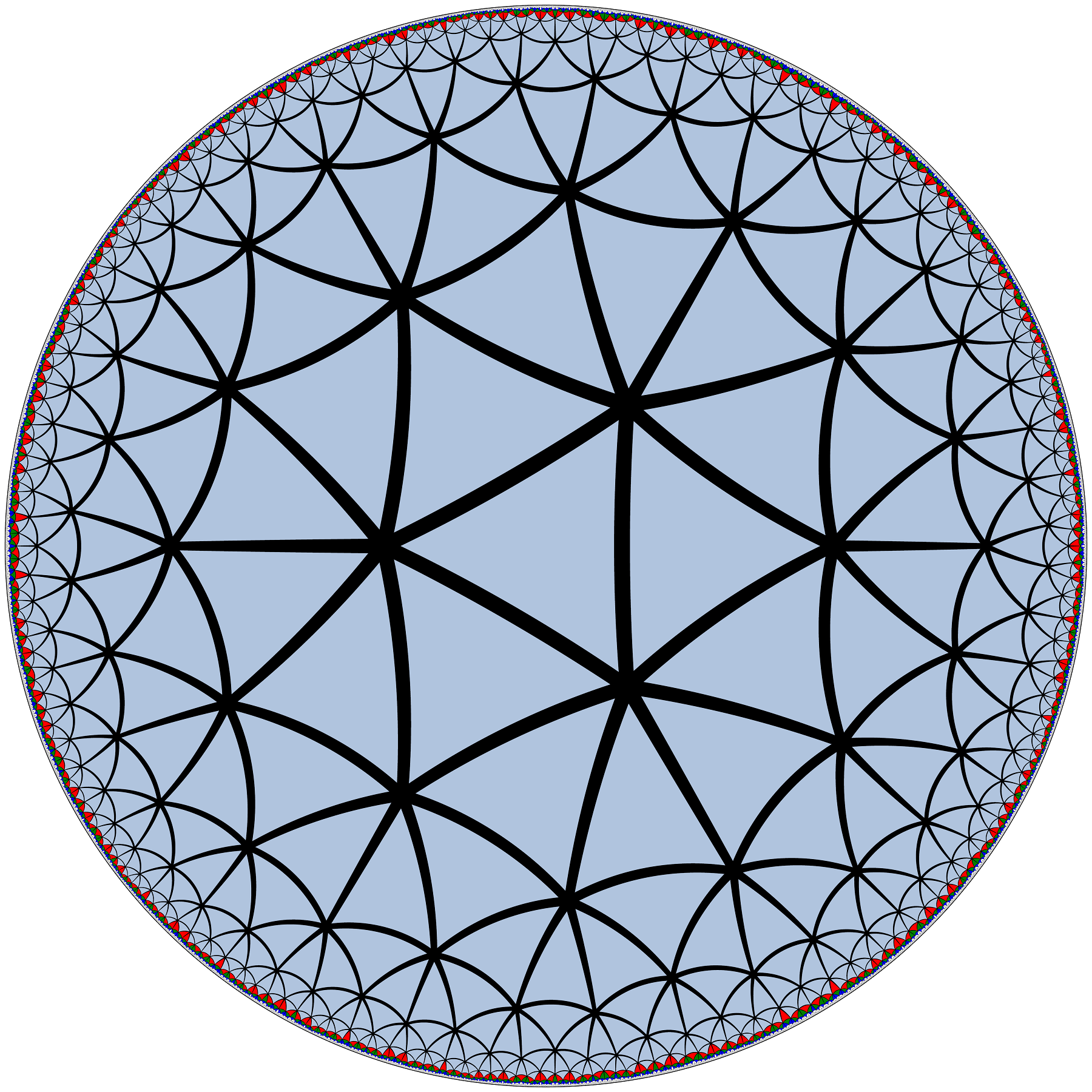}%
    }\vfill
    \subfloat[Boundary structure of the disk]{%
        \includegraphics[width=0.8\linewidth]{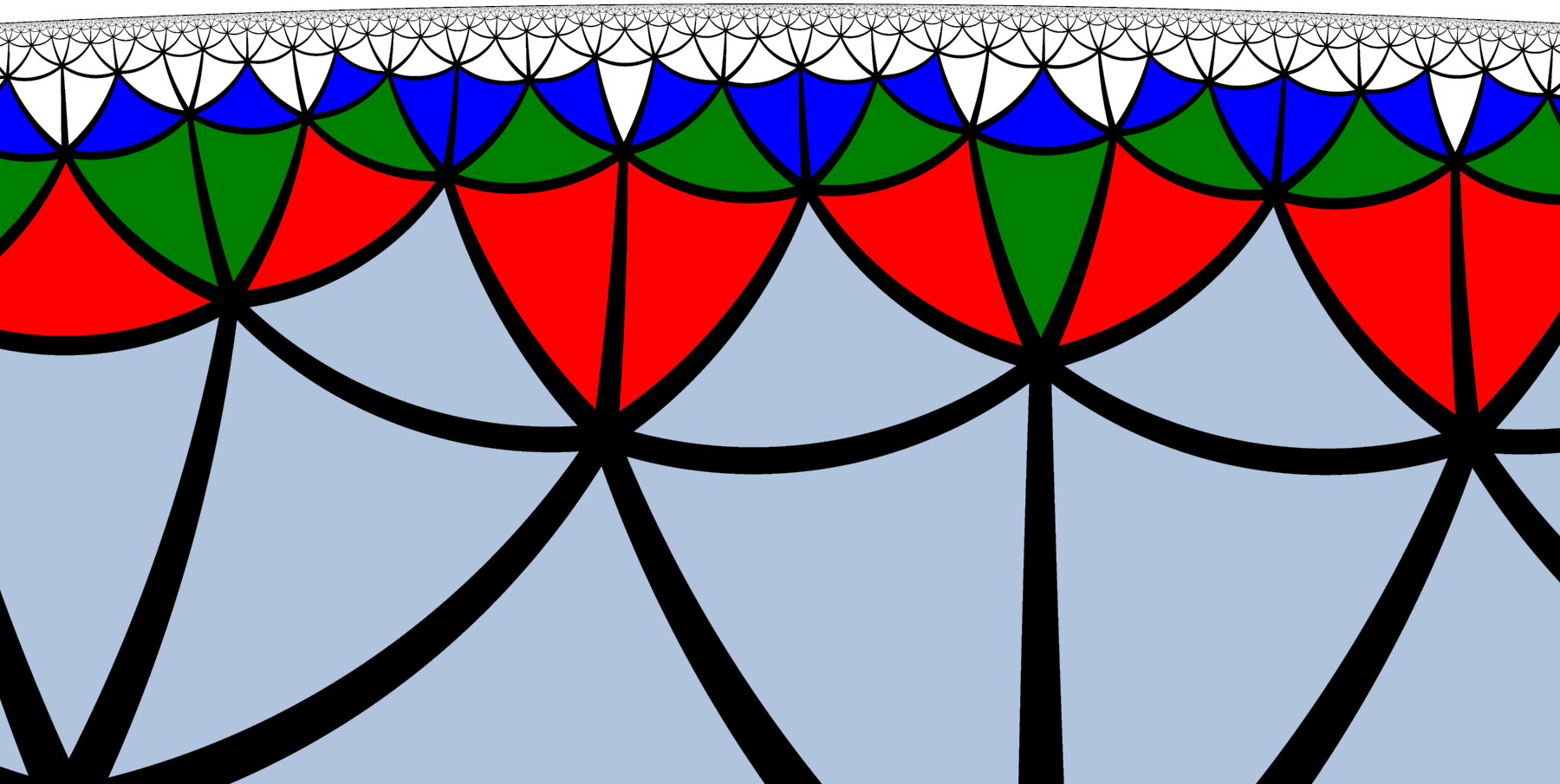}%
    }\vfill
\caption{(a) The full $\{3,7\}$ lattice, highlighting 10 layers and $591$ nodes. The first $10$ layers are drawn in one color and the subsequent $3$ layers each colored uniquely. (b) A close-up of the boundary of the lattice. Connectivity of the boundary vertices has no fixed pattern as additional layers are added.}
\label{lattice_layers} 
\end{figure}
\begin{figure}[!htpb]
\includegraphics[width=0.45\textwidth]{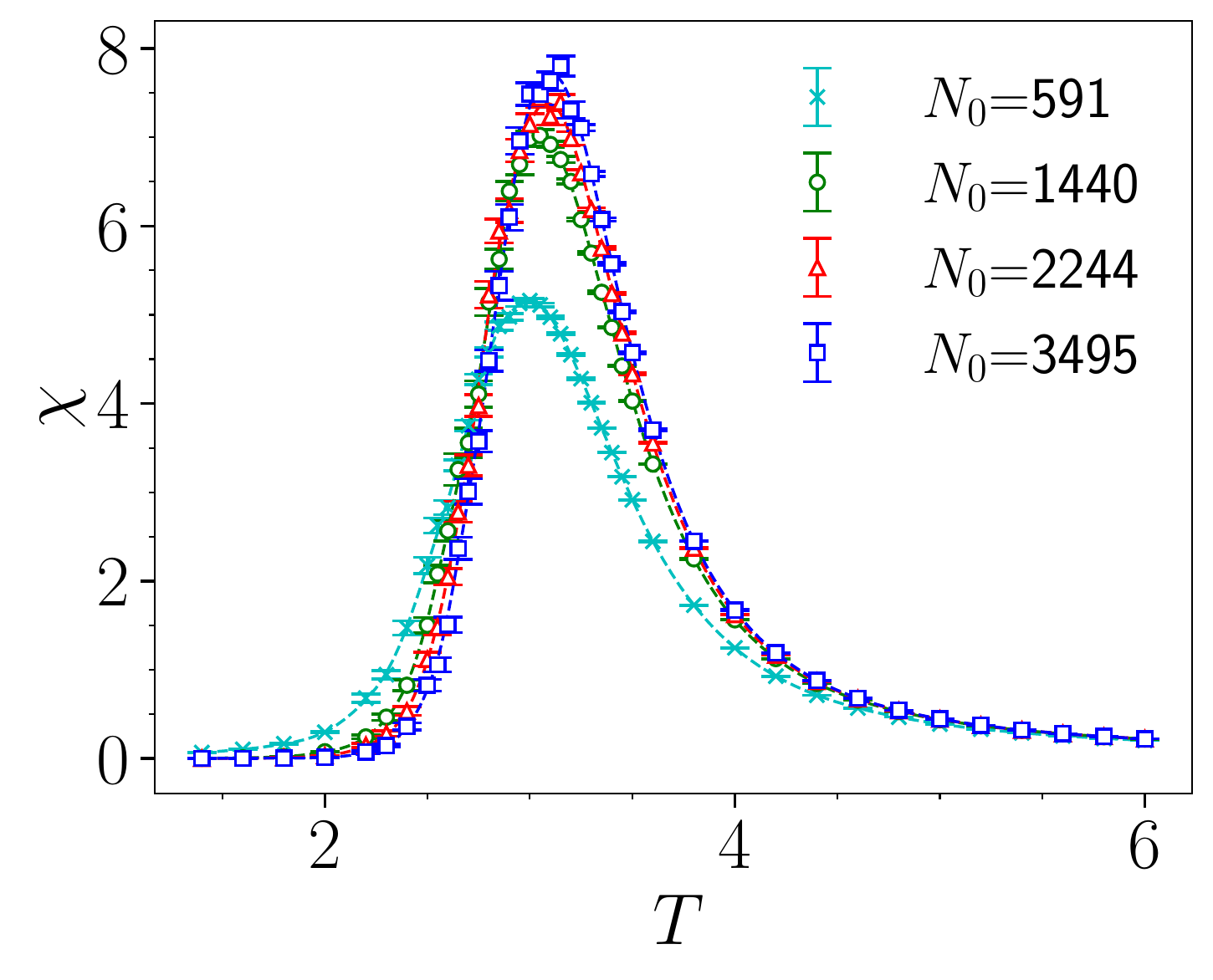}%

\caption{The bulk magnetic susceptibility computed from $N_{\mathrm{bulk}}=591$ spins
on the $10$ innermost layers of the tessellation as additional outer layers are added
up to a maximum of $N_{0}=3495$.}
\label{bulksus}
\end{figure}
\begin{figure}[!htbp]
\includegraphics[width=0.45\textwidth]{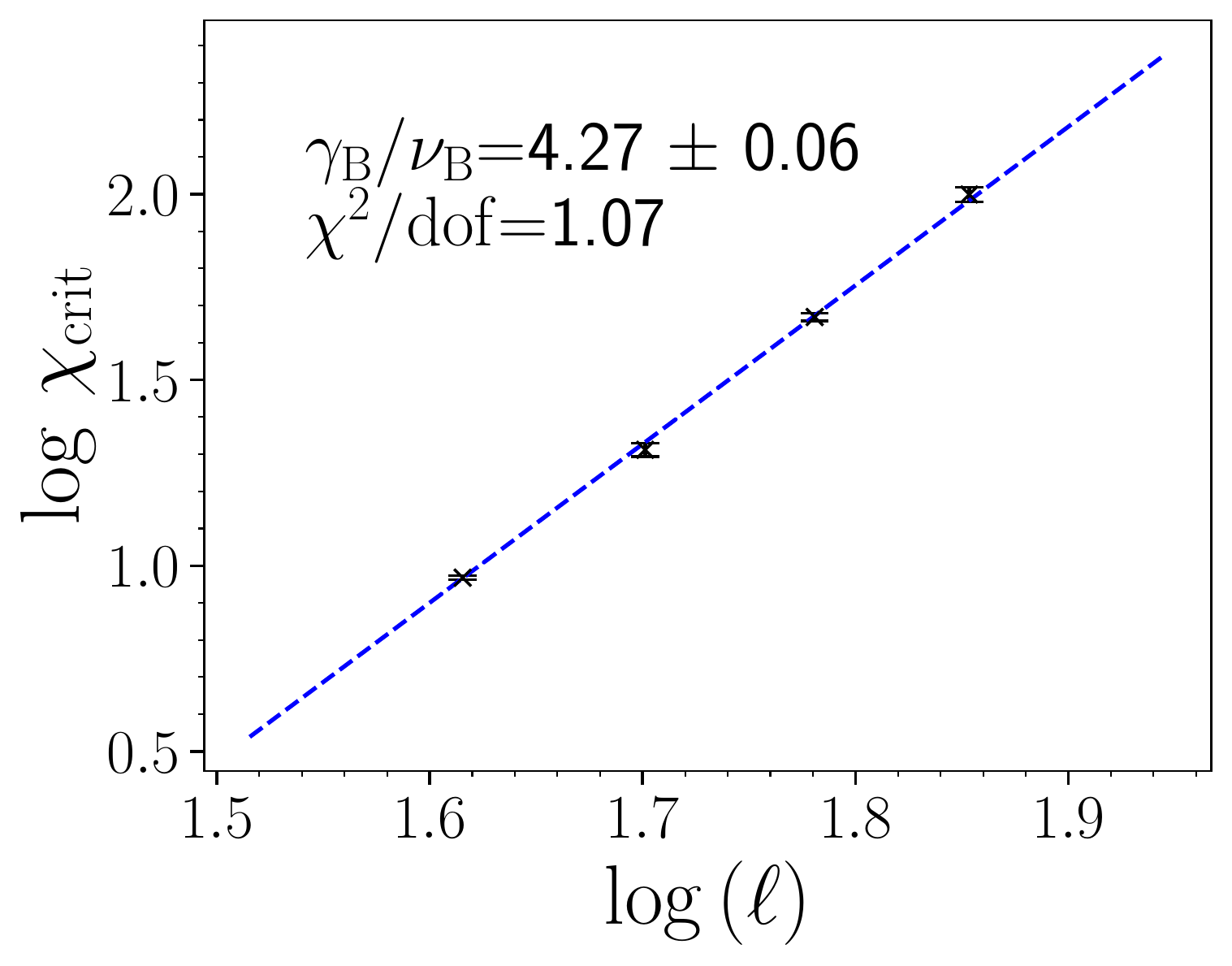}
\caption{The peak in the bulk susceptibility vs $\ell=\log(N_{\mathrm{bulk}})$ in log-log coordinates. The scaling exponent $\gamma_{\rm B}/\nu_{\rm B}$ corresponds to the slope of the fitted line.}
\label{scaling_bulk_sus}
\end{figure}

We construct a $\{3,7\}$ tessellated disk as shown in Fig.~\ref{lattice_layers} to obtain a lattice representation of hyperbolic geometry.\footnote{Regular two-dimensional hyperbolic geometry can be represented by Schl\"{a}fli symbol, $\{p,q\}$ where $p$ represents the $p$-sided polygon used as a building block to create the lattice and $q$ represents the coordination number of each vertex} Details of the lattice construction can be found in Ref.~\cite{Asaduzzaman:2020hjl}. The use of the triangle  symmetry group for the lattice construction was also emphasized by Brower \etal in Ref.~\cite{brower2019lattice}.
We place Ising spins on each vertex of this tessellation. The partition function of the nearest-neighbor Ising system is then given by
  \begin{equation} 
      Z= \sum_{\{ s \} } \Bigg[ \prod_{\langle ij\rangle} \exp (\beta  s_i s_j) \Bigg].
  \end{equation}
where, $\beta$ is the inverse of the temperature ($T$), the product $\prod_{\langle ij \rangle}$ is over all nearest-neighbor pairs, and the sum $\sum_{\{s\}}$ is over all possible spin configurations. We simulate the model with Markov chain Monte Carlo using the Metropolis and Wolff cluster algorithms \cite{wolffCollectiveMonteCarlo1989}.

We use open boundary conditions in our work. Since the fraction of vertices on the boundary relative to the bulk is essentially constant as the volume increases for such a tessellation, one must then be careful in defining the expectation value of bulk observables. We have examined the dependence of such bulk expectation values on the distance from the boundary by comparing expectation values over a fixed set of innermost layers of the tessellation as additional outer layers are added. Figure~\ref{bulksus} shows a plot of the bulk magnetic susceptibility
versus temperature for a series of lattices ranging up to $N_{0}=3495$ vertices where the bulk quantity is only computed using the $n=10$ innermost layers corresponding to $N_{\mathrm{bulk}}=591$ spins.\footnote{Other bulk thermodynamic quantities are shown in appendix~\ref{App1}} It is straightforward to extrapolate such data to the case where the bulk lies an infinite distance from the boundary. In practice, we observe that allowing for three outer layers leads to results that are independent of these limiting values within statistical errors.

Using simulations with just these three additional outermost layers allows us
to examine the finite-size scaling of bulk quantities.
For example, the scaling of the
peak of the bulk susceptibility is shown in Fig.~\ref{scaling_bulk_sus} which plots the logarithm of
the peak height versus the logarithm of the linear scale $\ell$ that characterizes the geometry. On a hyperbolic disk, it is
natural to use $\ell=\log{N_{\rm bulk}}$ since this characterizes bulk lattice geodesics. The system is pseudocritical once the correlation length approaches this scale. The slope of the linear fit yields an estimate of the bulk critical exponent $\gamma_{\rm B} / \nu_{\rm B} = 4.27(6)$.

Notice that the use of an open boundary
condition means that our current work differs 
from earlier studies. For example, Ref.~\cite{breuckmannCriticalPropertiesIsing2020} studies the bulk properties of the model having imposed a periodic boundary condition on the boundary. This 
boundary condition corresponds to using a hyperbolic manifold with genus $g \geq 1$ \cite{breuckmannCriticalPropertiesIsing2020,saussetPeriodicBoundaryConditions2007}. In contrast, Nishino \etal use a fixed ferromagnetic boundary condition in their Corner Transfer Matrix renormalization group approach \cite{Krmr2008IsingMO}. Earlier works using Pad\'{e} approximations from the low and high-temperature expansion can be found in Ref.~\cite{rietmanIsingModelHyperlattices1992}.
  
However, the open boundary condition we employ in the current work is the more natural choice in a holographic
context that closely resembles a Dirichlet condition. The scheme we use for implementing the open boundary condition is similar in spirit to the work of Shima \etal~\cite{Shima:2005vq}. 

\section{Boundary Thermodynamics}\label{boundary}
\begin{figure*}
    \hspace{5pt}
    \subfloat[\label{boundary_m}]{%
      \includegraphics[height=.25\textheight]{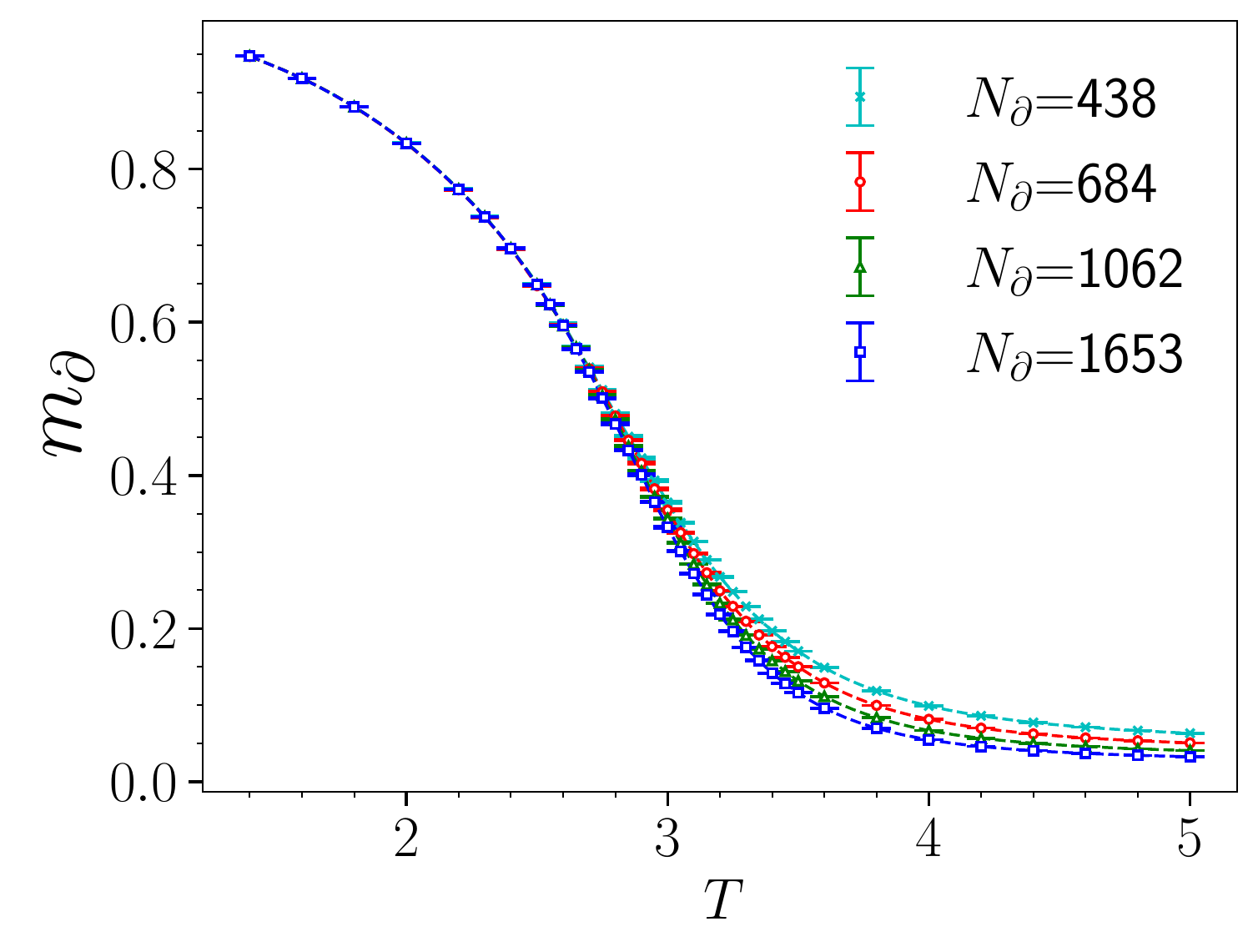}%
    }\hfill
    \subfloat[\label{boundary_sus}]{%
      \includegraphics[height=.25\textheight]{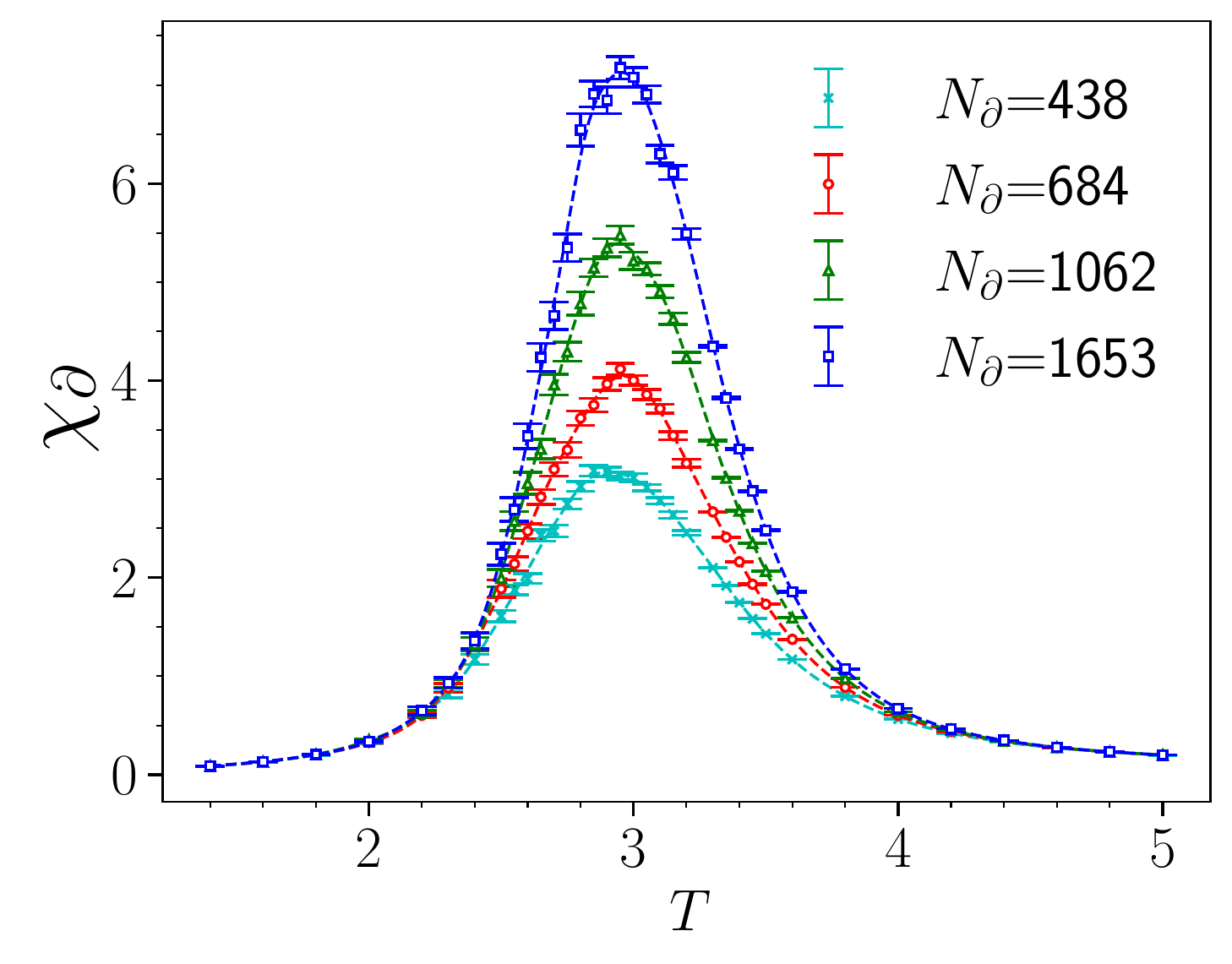}%
    }   \hspace{20pt}
    \caption{(a) Boundary magnetization and (b) boundary susceptibility are plotted against bulk-temperature from 10 to 13 layered Poincar{\'e} disk with increasing number of boundary spins $N_{\partial}$.}
    \label{boundary_obs} 
\end{figure*}

\begin{figure*}[!t]
\subfloat[\label{gn2.7}]{%
  \includegraphics[height=.18\textheight]{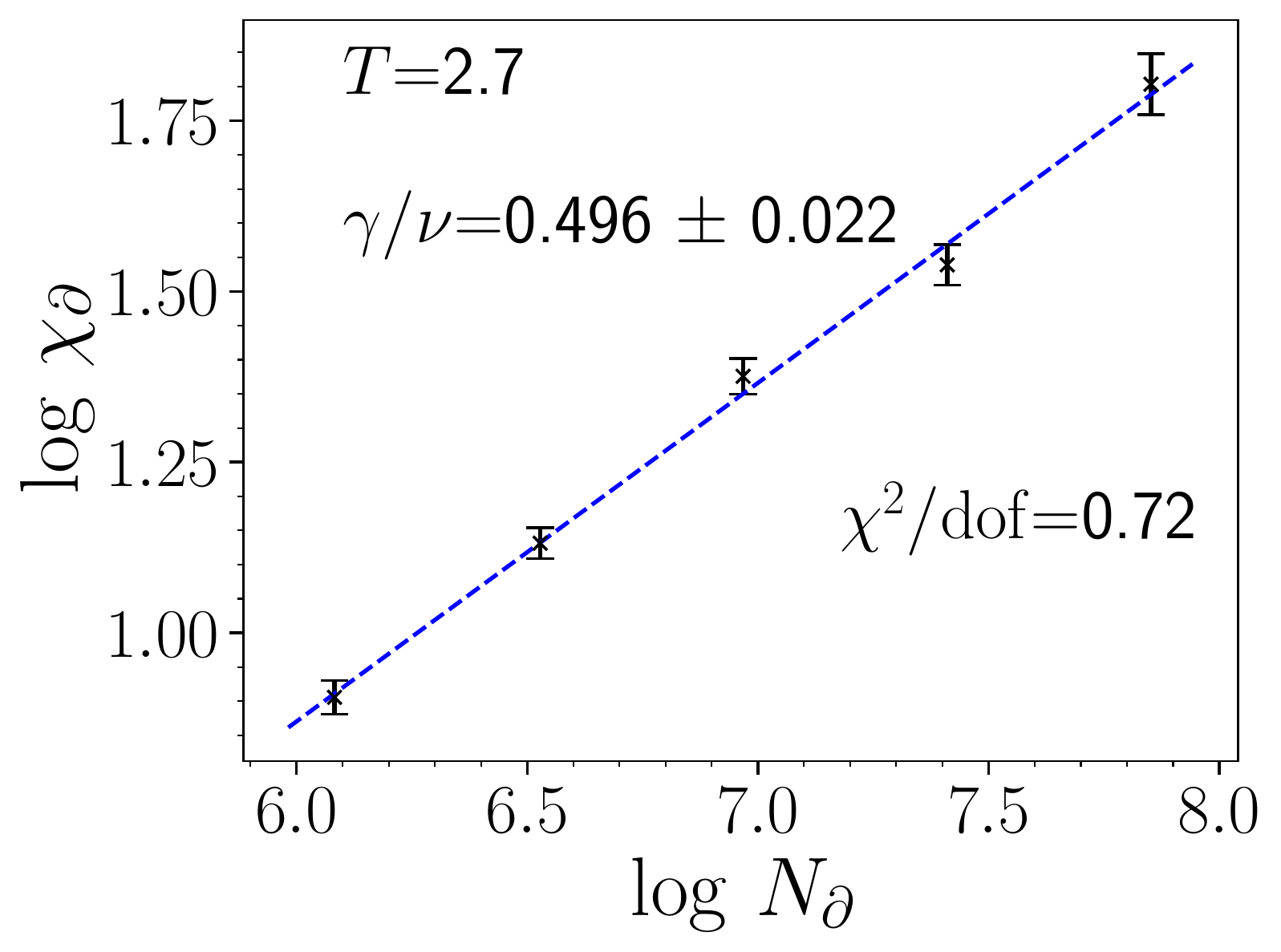}%
}\hfill
\subfloat[\label{gn3.0}]{%
  \includegraphics[height=.18\textheight]{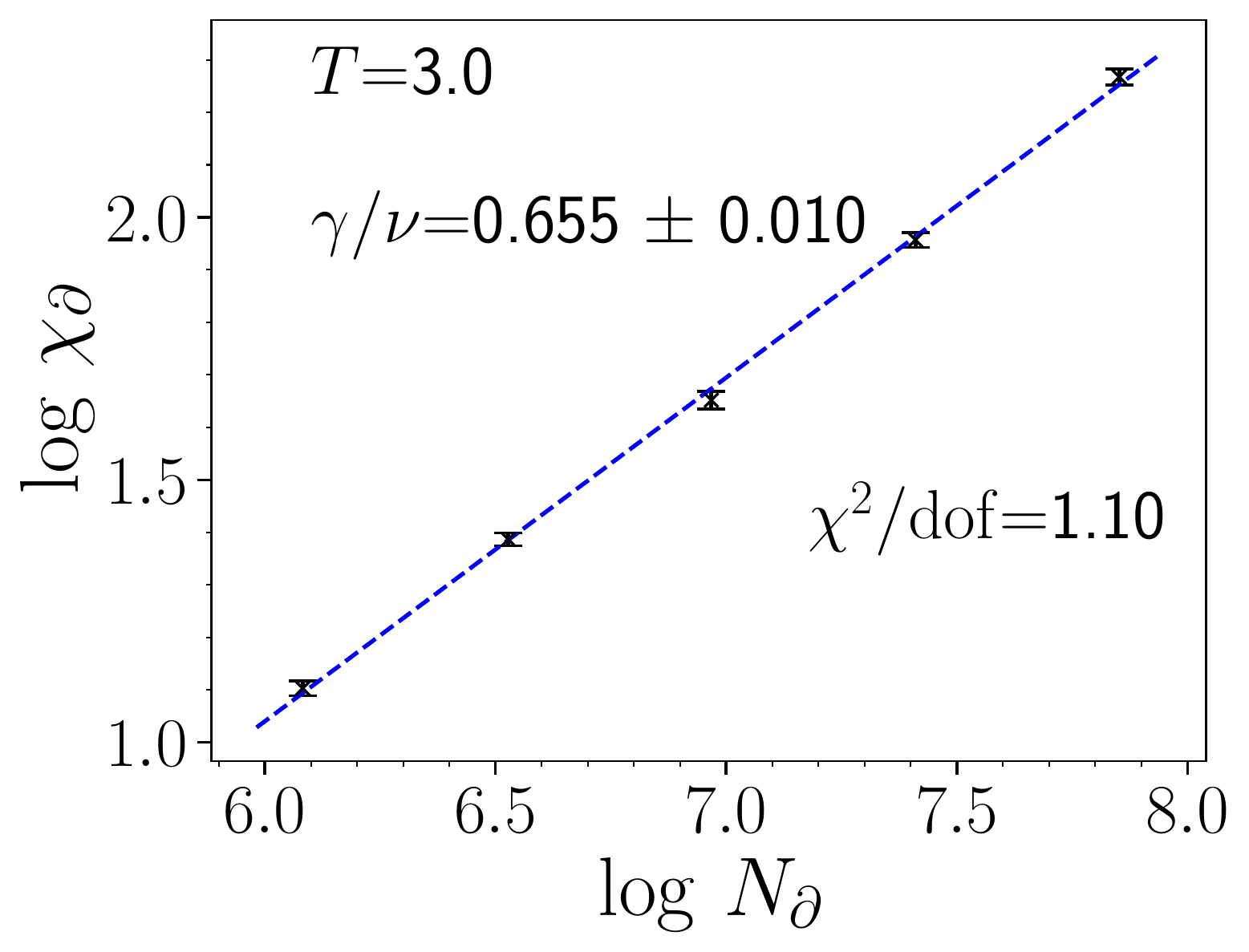}%
}\hfill
\subfloat[\label{gn3.4}]{%
  \includegraphics[height=.18\textheight]{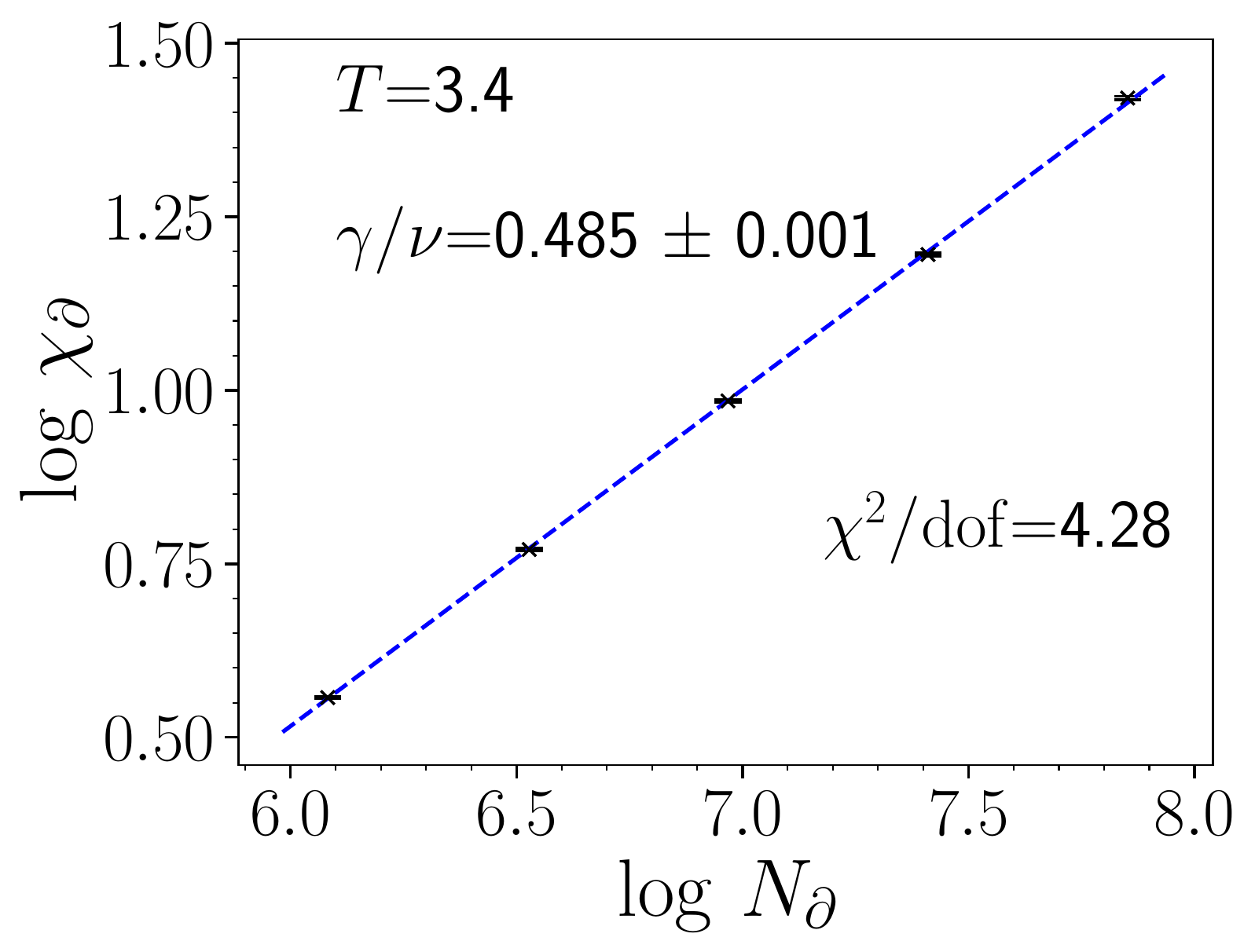}%
 }\
\caption{Logarithm of the boundary susceptibility ($\chi_\partial$) is plotted against logarithm of the total number of boundary points ($N_\partial$) for $T =$ (a) 2.7, (b) 3.0, (c) 3.4. Scaling exponents of the boundary susceptibility ($\gamma/\nu$) are computed from the linear fit of the data. }
\label{low_mid_high_scaling}
\end{figure*}

In this section, we will focus on the boundary observables. Fig.~\ref{boundary_m} and Fig.~\ref{boundary_sus} show the plot of (the absolute value of) the boundary magnetization and magnetic susceptibility versus temperature for a range of lattice sizes. Notice that the latter exhibits a peak close to that seen in the bulk susceptibility. However, this peak is broad and shows no sign of narrowing with increasing lattice size. Indeed, if we attempt a finite-size scaling analysis of the susceptibility we find evidence for a line of a continuously varying critical exponent $\gamma / \nu$---see Fig.~\ref{low_mid_high_scaling}.

\begin{figure*}[!t]
\subfloat[\label{corr2.7}]{%
  \includegraphics[height=.18\textheight]{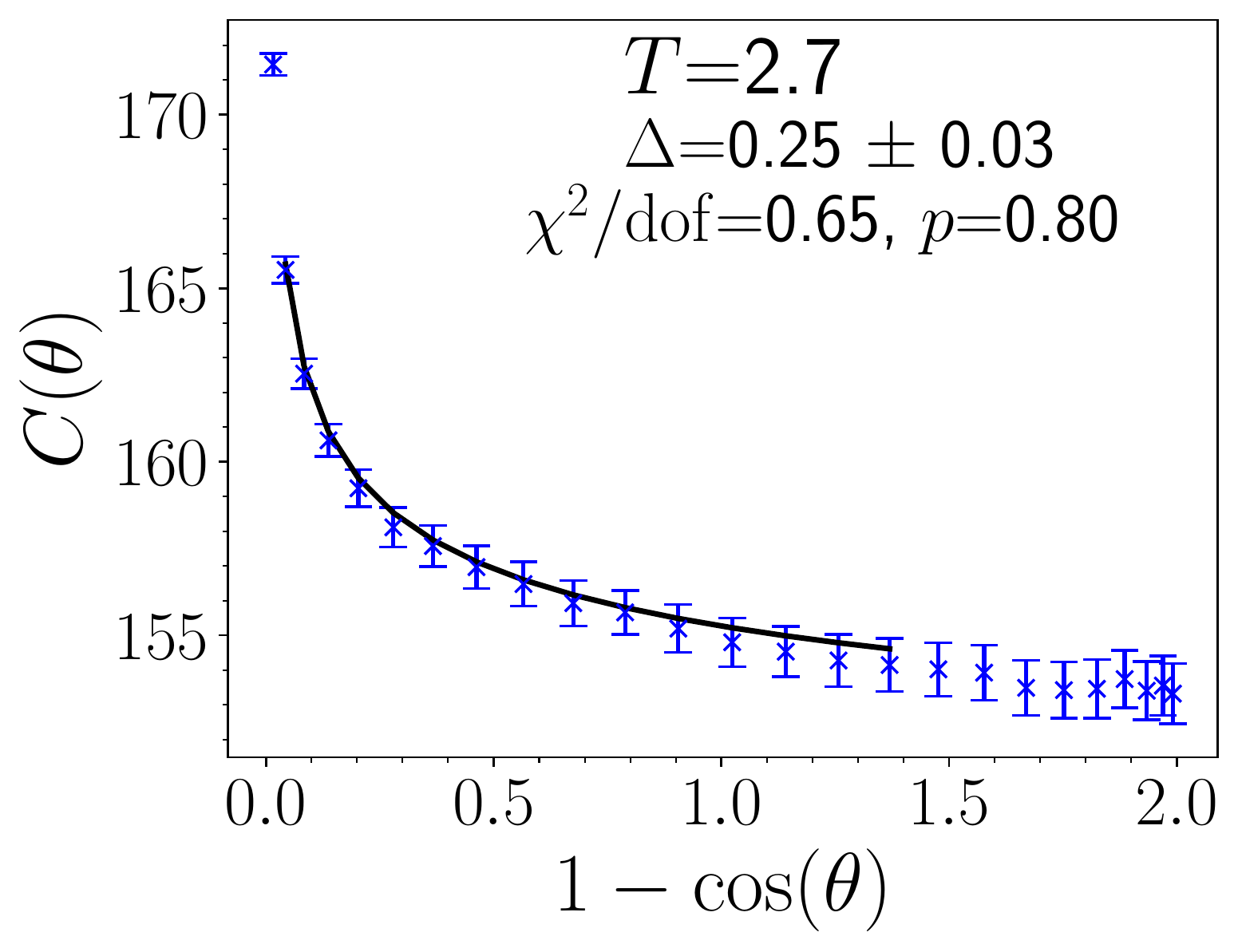}%
}\hfill
\subfloat[\label{corr3.0}]{%
  \includegraphics[height=.18\textheight]{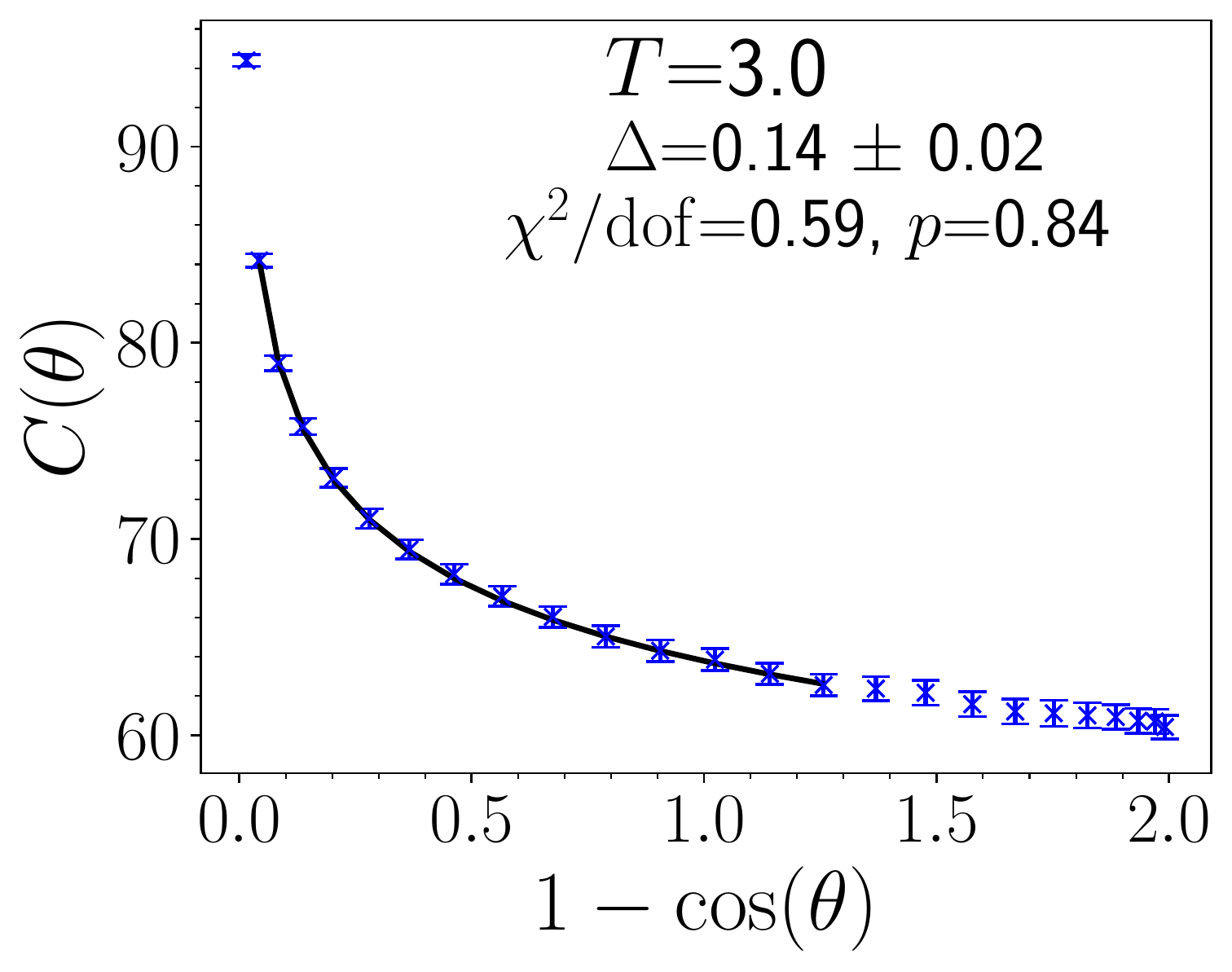}%
}\hfill
\subfloat[\label{corr3.4}]{%
  \includegraphics[height=.18\textheight]{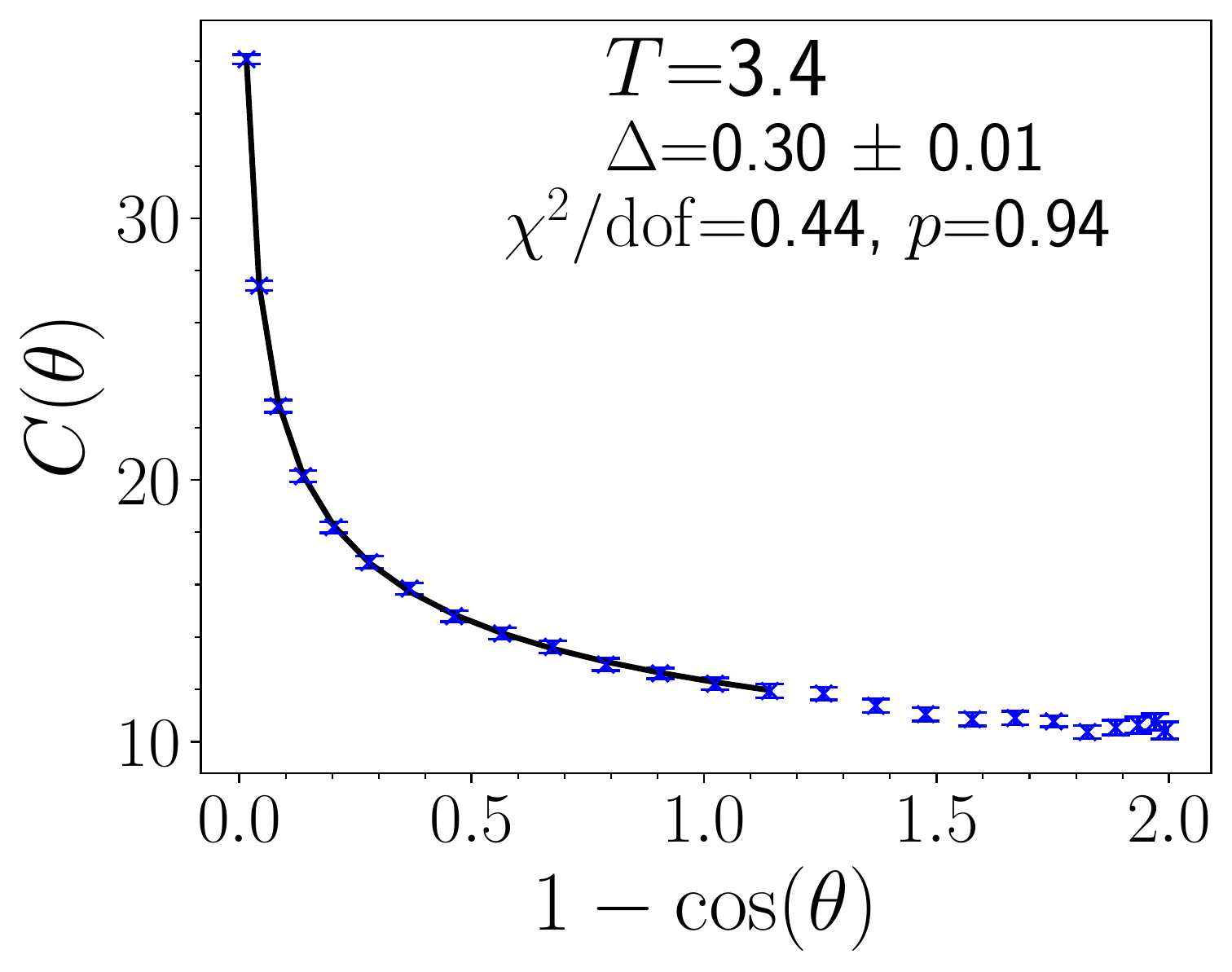}%
 }\
\caption{The boundary-boundary correlation function at $T = $ (a) 2.7, (b) 3.0, (c) 3.4, plotted against boundary distance squared $r^2\sim (1-\cos \theta)$. Results shown here are from the analysis of a 12 layered Poincar{\'e} disk with boundary length $N_\partial=1062$. $\chi^2$ per degree of freedom and the p-value of the fits are noted in the figures.}
\label{low_mid_high_corr}
\end{figure*}

Associated with this scaling, we can examine the boundary-boundary correlators over the same range of temperatures.  Correlation functions for three temperatures are shown in Fig.~\ref{low_mid_high_corr} with a choice of ``best-fit'' with the solid line. After binning the data, we proceed using a single-elimination jackknife.  Then, we perform correlated fits.  We fit over several fit ranges to estimate a systematic error associated with our ``best'' choice of fit range.  
Our fit ansatz has the form,
\begin{equation}
    \langle s(0) s(r) \rangle = a(T)+b(T)r^{-2\Delta(T)},\label{eqn_fit}
\end{equation}
where the distance measured on the boundary $r$ can be traded for an angle via the relation $r^2\sim \left(1-\cos\theta\right)$. Notice that the conformal behavior is given by the connected correlator which is insensitive to $a(T)$.
The final error shown in the plots includes the systematic error associated with the fit range added in quadrature with the statistical error obtained from the best fit. 

\begin{figure}[!htb]
  \includegraphics[width=\linewidth]{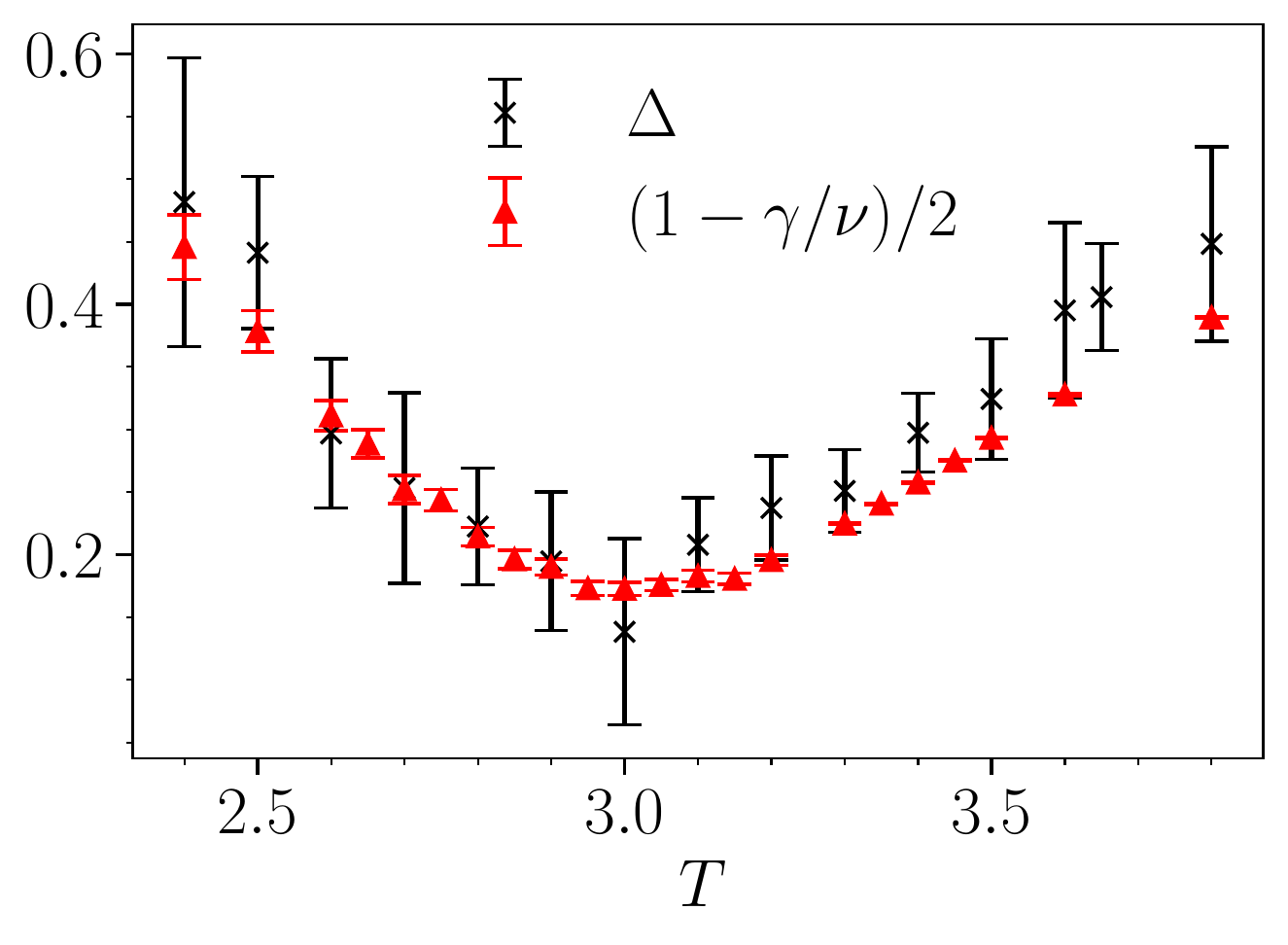}%
 \caption{The scaling exponent of the boundary spin operator computed from fits to the boundary two-point correlator, denoted by $\Delta$, and from the finite-size scaling of the boundary susceptibility, denoted by $(1-\gamma/\nu)/2$.}\label{Delta_vs_T}
\end{figure}

The boundary susceptibility is of course nothing more than the integral of this correlation function, and hence we predict that the susceptibility exponent $\gamma / \nu = 1-2\Delta$. A plot showing the value of the scaling dimension extracted from the susceptibility together with the value obtained by a power-law fit to the correlation function is shown in Fig.~\ref{Delta_vs_T}. The agreement is excellent, furnishes a nice consistency test of our procedure, and provides strong evidence that the boundary theory indeed exhibits power-law behavior both at high and low temperatures. Notice that once $1-2\Delta <0$, the susceptibility no longer diverges with lattice size, which explains the location of the edges of the broad peak shown in Fig.~\ref{boundary_sus}.

In fact it is easy to see that the boundary correlation function should
exhibit a power law for high temperature. Expanding the Boltzmann factors for
small $\beta$ we find the correlator is given by
\begin{equation}
\label{eq:high-temp-spin-spin}
    \langle s_k s_\ell \rangle  \propto \sum_{ \{ s \} } s_k s_\ell  \Big( \prod_{\langle ij \rangle}( 1 + s_i s_j \tanh \beta) \Big).
\end{equation}
The leading-order contribution in $\beta$ corresponds to the minimal-length path in the lattice between the two boundary spins. On a hyperbolic disk this path runs through the bulk and yields 
\begin{equation}
    \langle s_k s_\ell \rangle \propto (\tanh\beta)^R
\end{equation}
where $R$ is the length of the geodesic between $k$ and $\ell$. But $R\sim\log{r}$ on the hyperbolic  disk with $r$ the boundary distance between the spins. Thus we find
\begin{equation}
    \langle s_k s_\ell\rangle \propto r^{-2\Delta}
\end{equation}
where $\Delta\sim \frac{1}{2}\log{\coth\beta}\sim -\log\beta\;{\rm for}\;\beta\to 0$.
Thus it is natural to expect a conformal boundary phase for a range
of high temperatures. 

To understand the low-temperature behavior we can perform a duality transformation on the model \cite{savitDualityFieldTheory1980a}. This maps the original Ising  system with spins $\{s_i\}$ on a $\{3,7\}$ tessellation at (inverse) temperature $\beta$ into another Ising model with spins $\{\sigma_j\}$ living on the dual $\{7,3\}$ tessellation at temperature $\widetilde{\beta}=\frac{1}{2}\log{\coth{\beta}}$. Notice that high temperatures in the original model are mapped to low temperatures in the dual model. Furthermore, the functional relationship between the boundary distance and the length of the corresponding bulk geodesic is the same for both $\{3,7\}$ and $\{7,3\}$ tessellations. This implies that the power-law behavior of the $\{s_i\}$ system at high temperature produces a power law correlator for the dual $\{\sigma_i\}$ system at low temperature. But the dual system is just another discretization of hyperbolic space and so one concludes that the $\{s_i\}$ boundary correlator on the original lattice should also possess power-law behavior at low temperature with $\Delta\sim \log \coth\widetilde{\beta}\sim \beta\;{\rm for}\;\beta\to\infty$. And indeed, this is precisely what is observed in our simulations.\footnote{See appendix~\ref{App2} for the correlator results of the dual lattice.} Notice that the minimal value of the boundary scaling dimension $\Delta$ is obtained for $T\sim T_c$ corresponding to the point where the bulk mass gap on hyperbolic space has been tuned to zero.

\section{Summary and Prospects}
\label{sec:summary}
We have simulated the Ising model on a tessellation of the hyperbolic disk, exploring the AdS/CFT correspondence at strong coupling on the lattice.  On a $\{3,7\}$ tesselation, we
find evidence for a bulk phase transition at $T_c\sim 3$ separating
a low temperature magnetized phase from a disordered phase at high temperature.  Since the number of boundary points is always a constant significant fraction of the bulk points, special care was taken to define bulk observables which are insensitive to the presence of this boundary.

The primary focus of this paper was the correlation functions and
thermodynamics of the Ising spins located at the
boundary. Our numerical results show that
boundary-boundary correlation functions exhibit power-law behavior over a wide
range of temperatures starting at high temperature, through the bulk phase transition, and persisting into the low-temperature ordered phase.
The high-temperature conformal phase is straightforward
to understand using high-temperature expansions and relies
only on the geodesic structure inherited by tessellations
of hyperbolic space. At low temperatures, arguments rooted in the duality transformation of the Ising model suggest that boundary criticality will persist in the ordered phase.

We conjecture that this low-temperature conformal boundary phase extends
all the way to $T=0$ in the thermodynamic limit but that the rapid increase in
the scaling  dimension as $T\to 0$ makes it difficult to verify this on
finite lattices. 

The fact that the boundary theory is scale-invariant at long distances for any temperature is consistent with the usual arguments for the AdS/CFT correspondence in the continuum which ties the conformal symmetry of the boundary theory to the bulk isometries of anti-de Sitter space. Since our tessellation approximates the latter and is fixed independent of temperature we would not expect the approximate conformality of the lattice theory to depend on couplings in the matter sector.

In this first study of its type, conformal invariance of the boundary theory was apparently maintained through a regime of strongly coupled bulk physics.  It will be fascinating to continue exploring the AdS/CFT correspondence in regions where perturbative methods are unavailable and lattice methods are the only tools in our arsenal.

\section{Acknowledgements} \label{sec:acknowledgements}
SC, JUY and MA would like to thank the QuLat collaboration and Rich Brower \& Cameron Cogburn in particular for stimulating discussions. MA would also like to thank Larne Pekowsky of the Information Technology department of Syracuse University for queries regarding running the jobs at the `Orange Grid'.
This work is supported in part by the U.S.\ Department of Energy (DOE), Office of Science, Office of High Energy Physics, under Award Number {DE-SC0009998}
and {DE-SC0019139}.  Numerical computations were performed at Syracuse University HTC Campus Grid: NSF award ACI-1341006.
This manuscript has been authored by Fermi Research Alliance, LLC under Contract No. DE-AC02-07CH11359 with the U.S. Department of Energy, Office of Science, Office of High Energy Physics.
\\


\appendix
\section{Other bulk observables\label{App1}}

\begin{figure*}[!htb]
\hfill
\subfloat[\label{bulkm}]{%
  \includegraphics[height=.18\textheight]{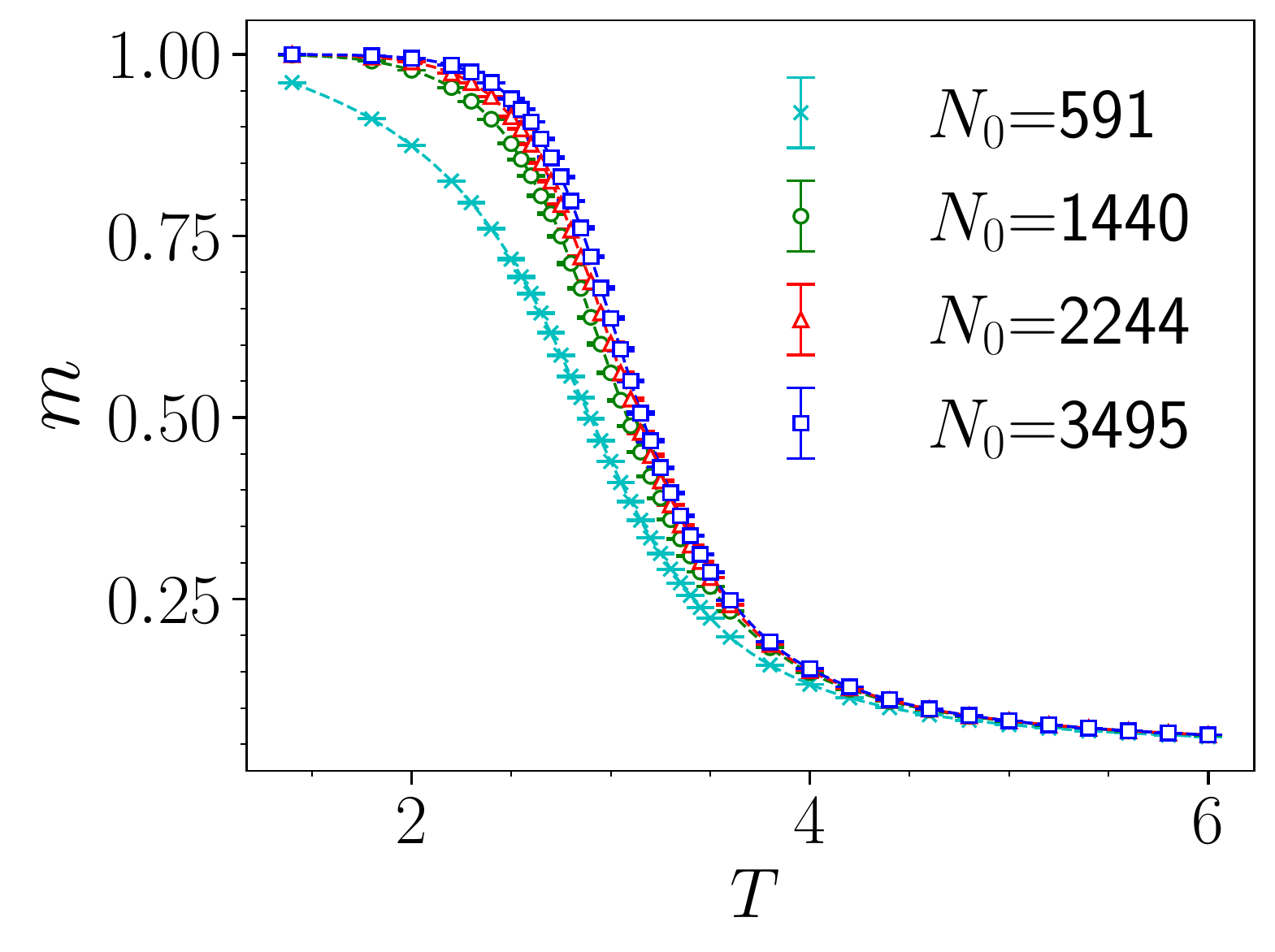}%
}\hfill
\subfloat[\label{bulke}]{%
  \includegraphics[height=.18\textheight]{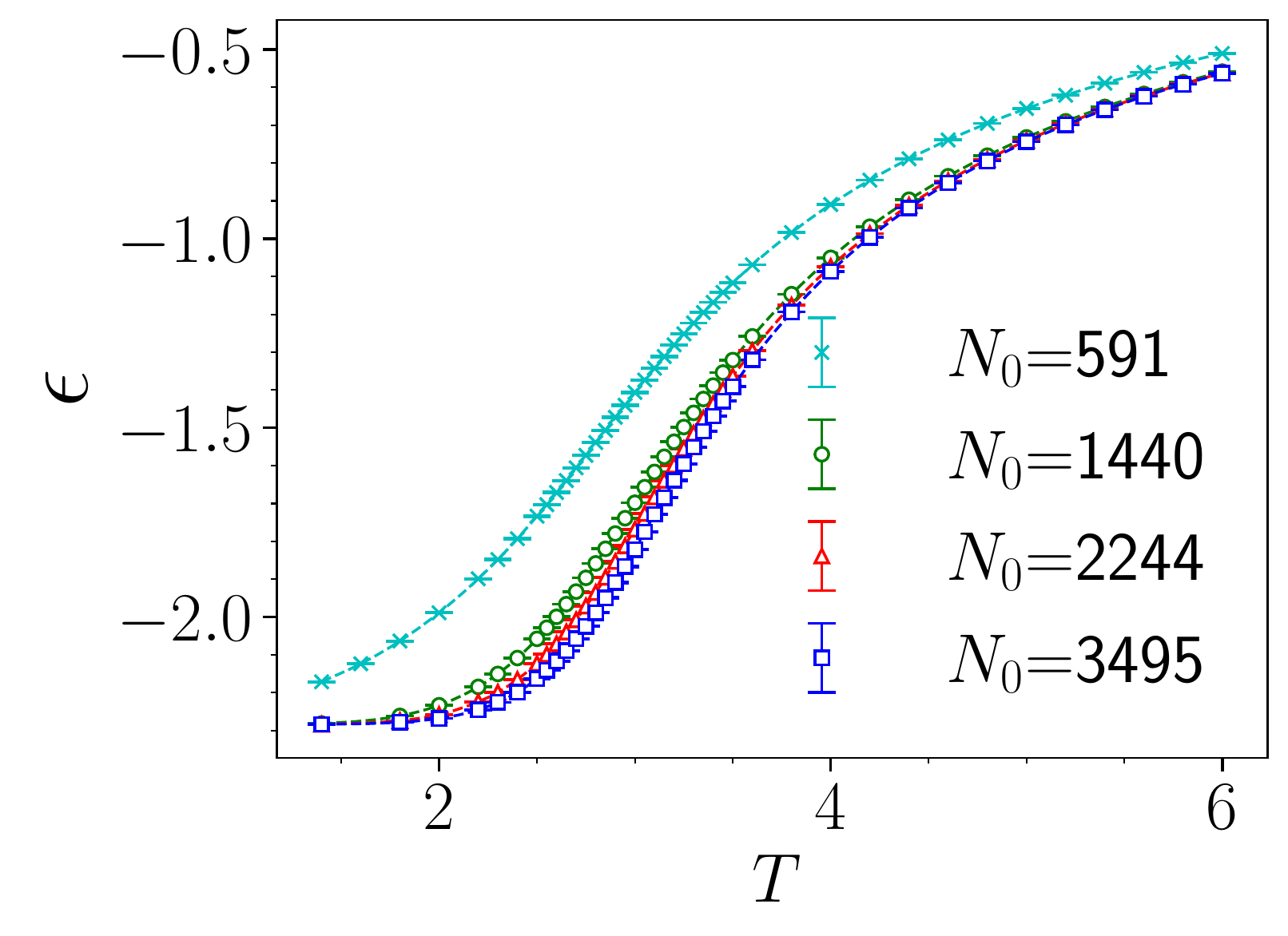}%
}\hfill
\subfloat[\label{bulkhc}]{%
  \includegraphics[height=.18\textheight]{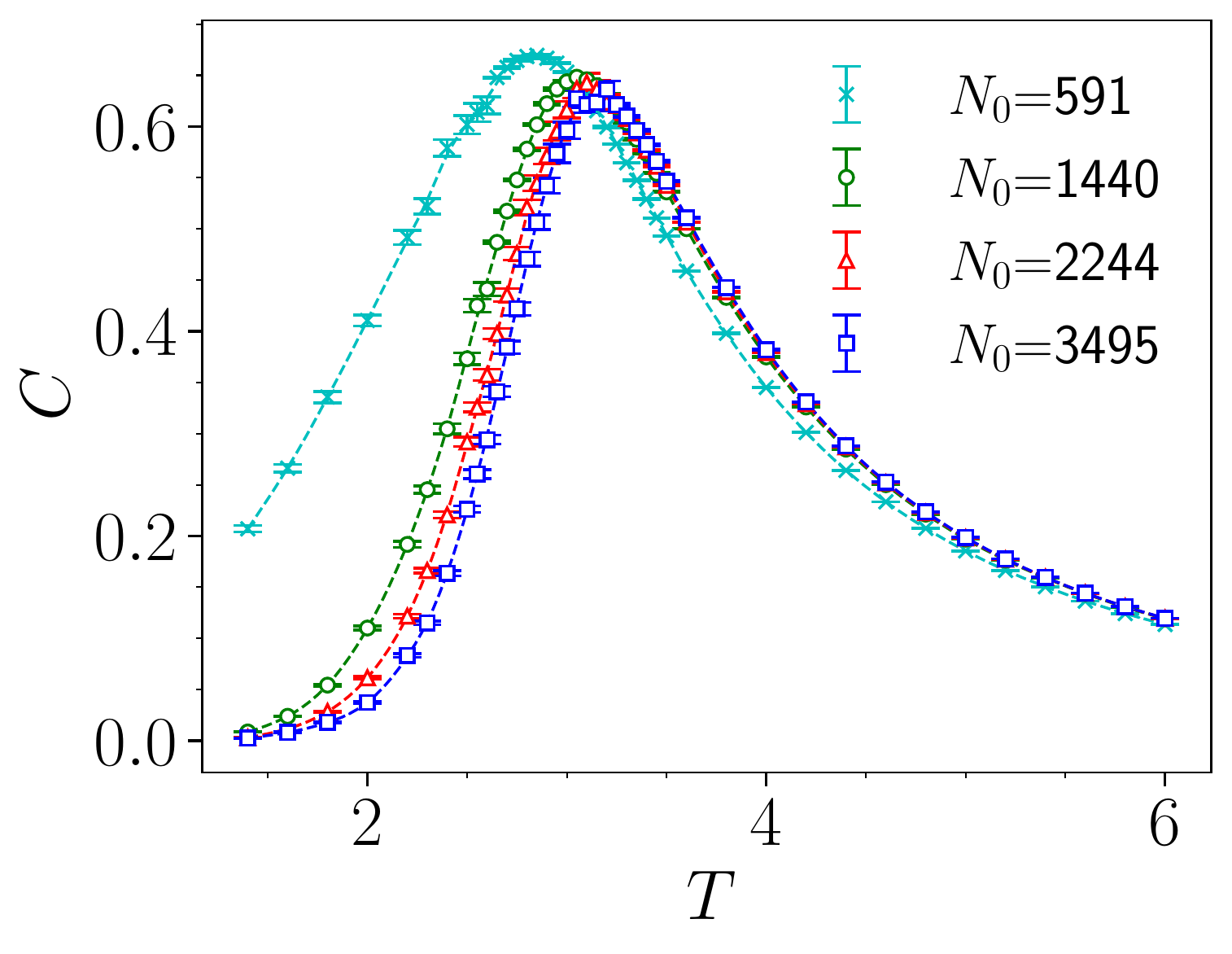}%
 }\hfill
\caption{(a) Bulk magnetization, (b) internal energy and  (c) heat capacity are computed from $N_{\mathrm{bulk}}=591$ spins
on $10$ innermost layers of the tessellation as additional outer layers are added up to a maximum of $N_{\mathrm{total}}=3495.$}
\label{bulkall}.
\end{figure*}


\begin{figure*}[!htb]
\hfill
\subfloat[\label{dcorr1.1}]{%
  \includegraphics[height=.17\textheight]{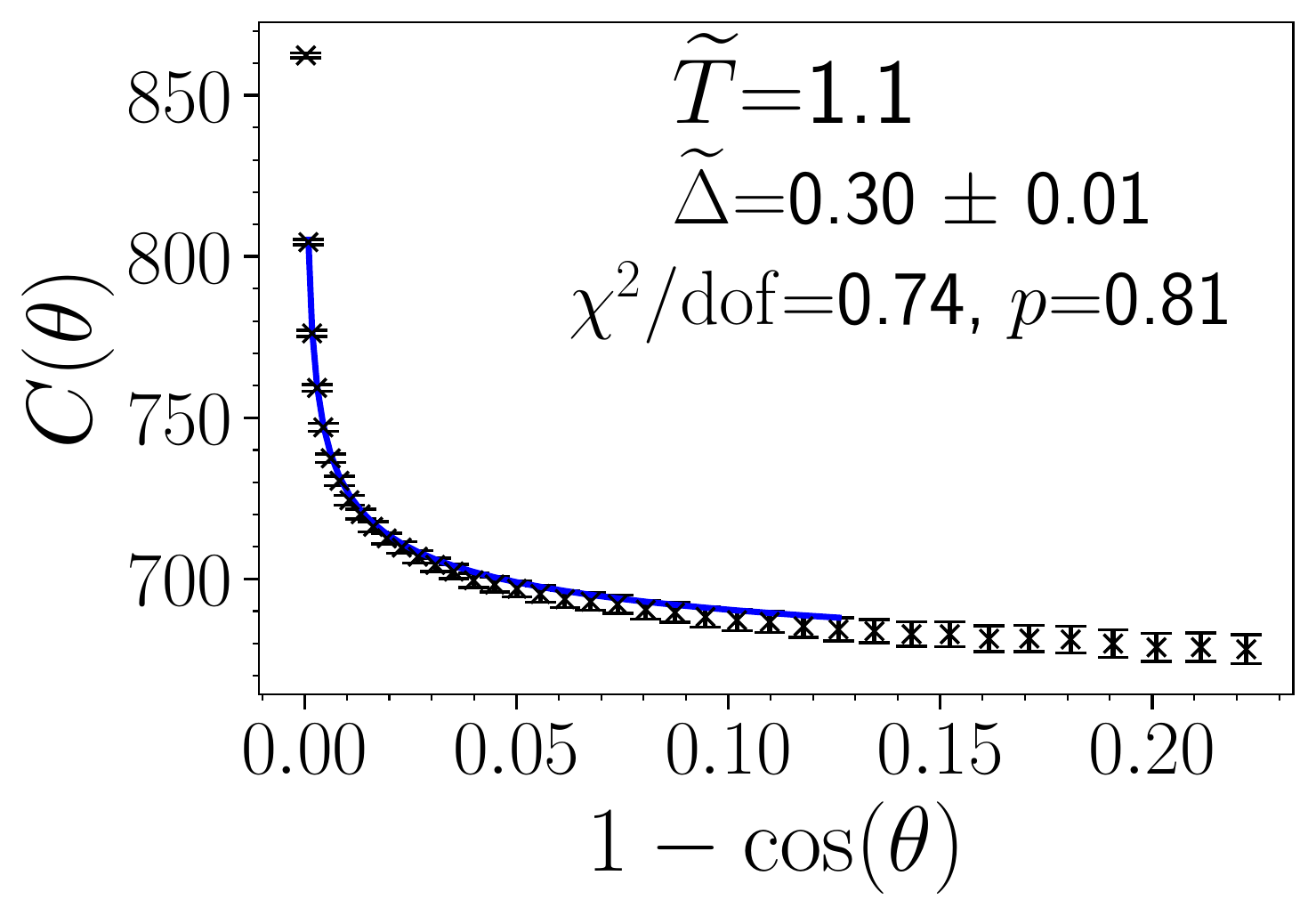}%
}\hfill
\subfloat[\label{dcorr1.2}]{%
  \includegraphics[height=.17\textheight]{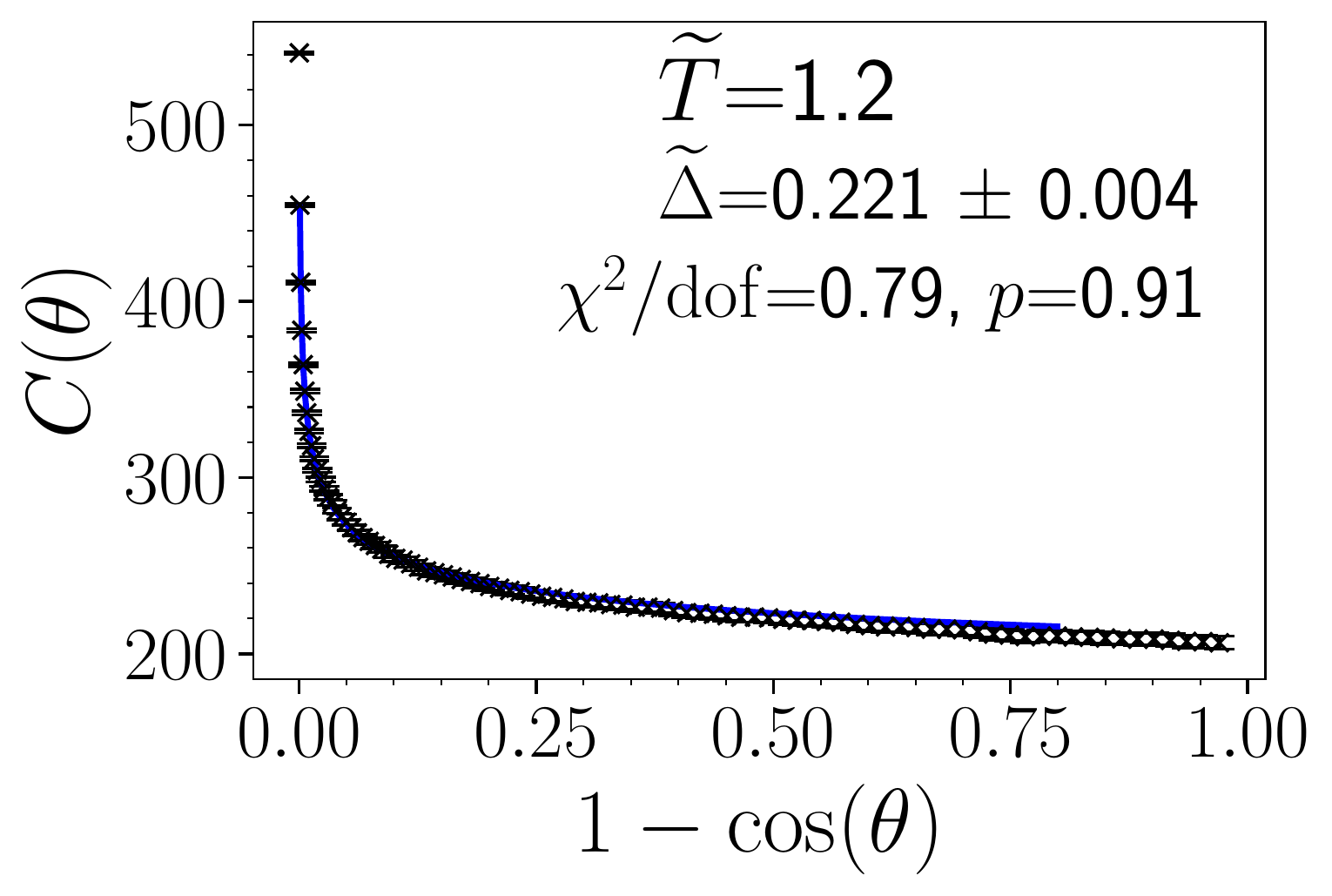}%
}\hfill
\subfloat[\label{dcorr1.3}]{%
  \includegraphics[height=.17\textheight]{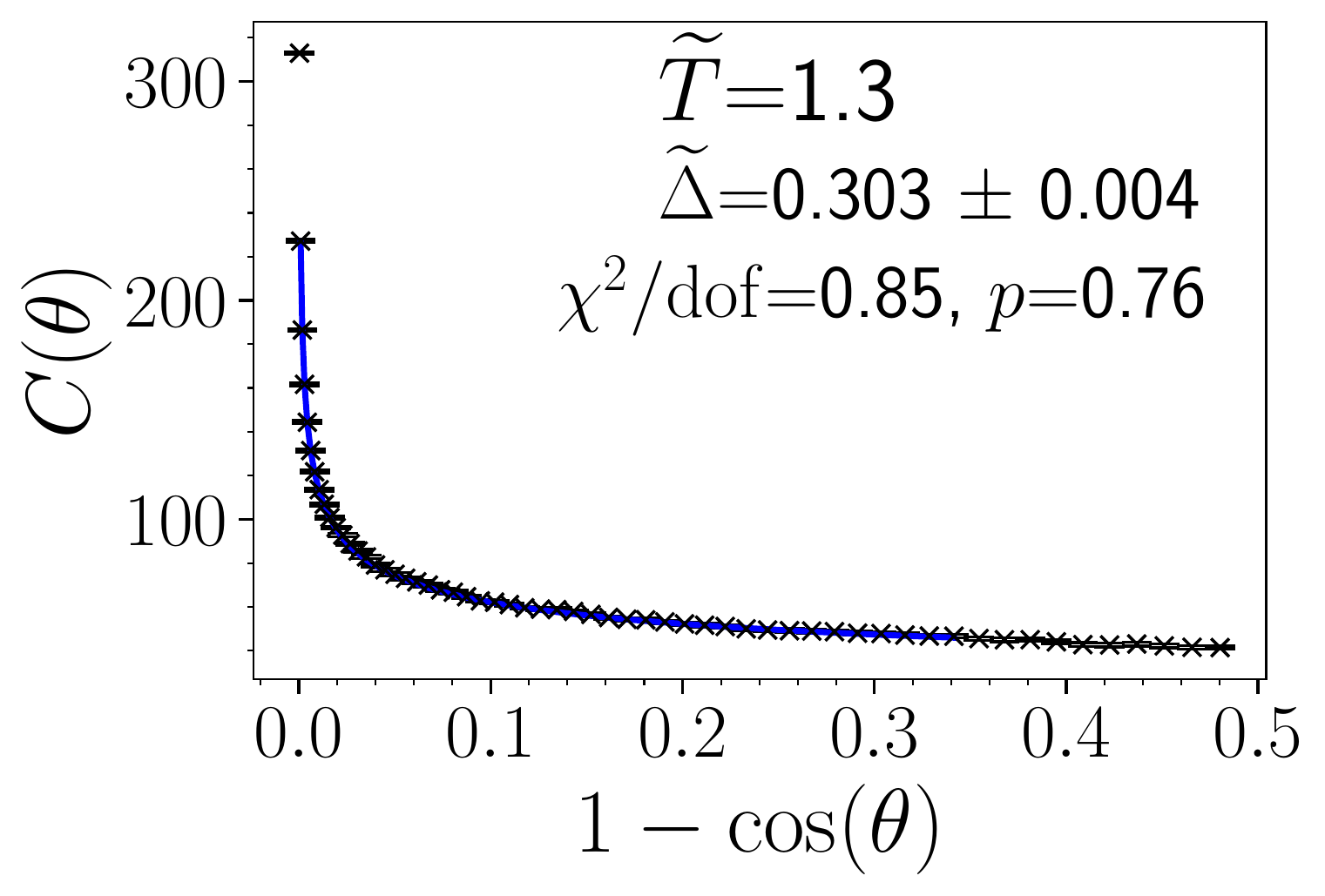}%
 }
 \hfill
\caption{The boundary-boundary correlation function of the dual spin variable ($C=\langle \sigma_0 \sigma_r \rangle$) at $\widetilde{T} = $ (a) 1.1, (b) 1.2, (c) 1.3 plotted against boundary distance squared $r^2\sim (1-\cos \theta)$. Results shown here are from the analysis of a six-layered $\{7,3\}$ Poincar{\'e} disk with boundary length $N_\partial=3647$. }
\label{dual_corr}
\end{figure*}

\begin{figure*}[!htb]
  \includegraphics[width=0.6\linewidth]{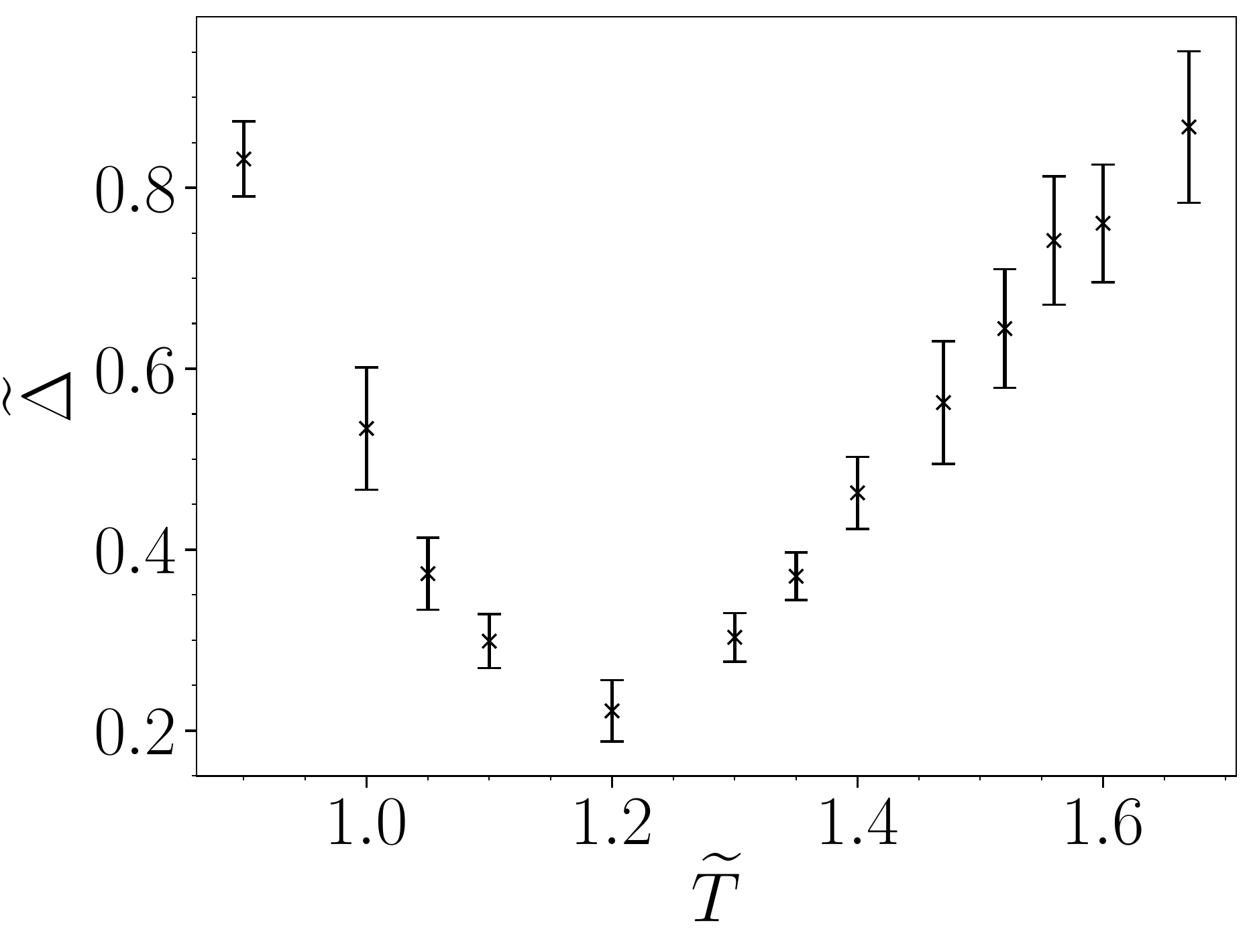}%
 \caption{Scaling exponent of the dual boundary spin operator ($\widetilde{\Delta}$) computed from the fits of the boundary correlator.}\label{dual_Delta}
\end{figure*}

\begin{figure*}[!htb]
\hfill
\subfloat[\label{delta_HT_fit}]{%
  \includegraphics[height=.25\textheight]{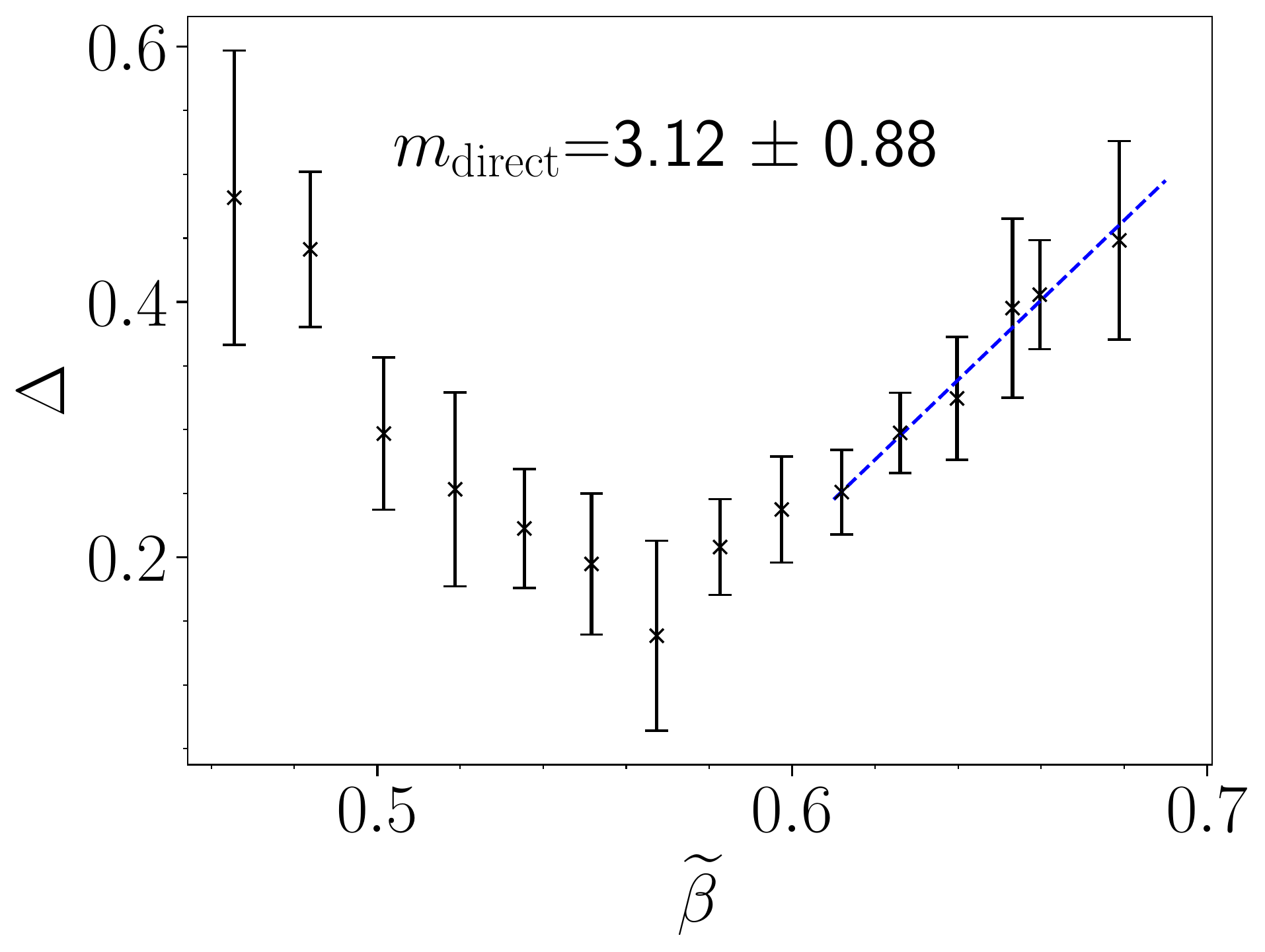}%
}\hfill
\subfloat[\label{dual_HT_fit}]{%
  \includegraphics[height=.25\textheight]{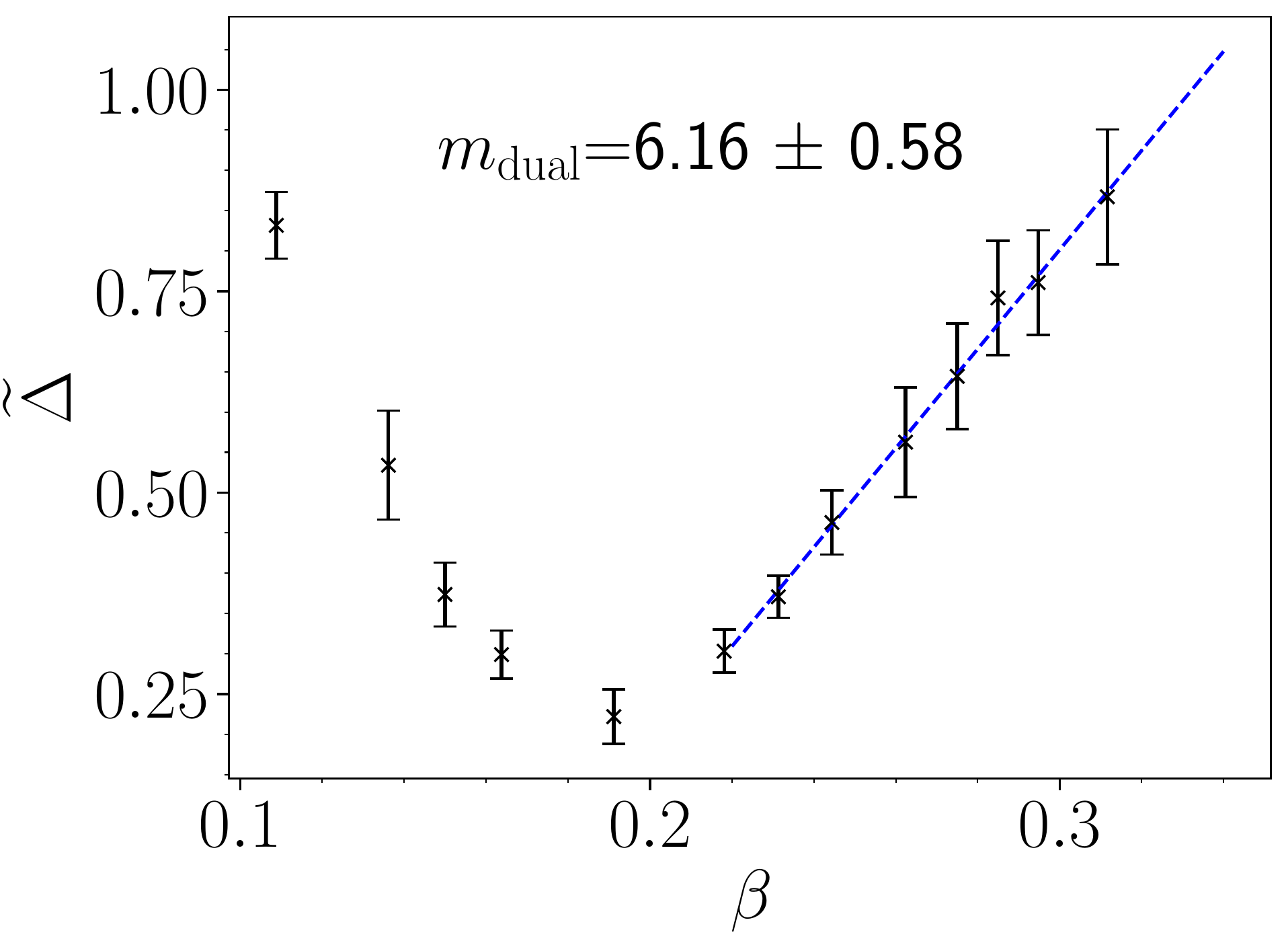}%
}\hfill 

\caption{(a) Scaling exponents at direct lattice $\Delta$ vs. dual inverse temperature ($\widetilde{\beta}$), and (b) scaling exponent at dual lattice ($\widetilde{\beta}$) vs. inverse temperature at direct lattice ($\beta$). Linear fits at high temperature are shown with the extracted slope denoted in the figure. }
\label{HighT_fit}
\end{figure*}

In this appendix, we present the results of the bulk observables like magnetization, energy and heat capacity computed from the Monte Carlo simulation. The following definitions were used to compute the absolute magnetization per spin 
\begin{equation}
     m=\frac{M}{N_{\mathrm{bulk}}} = \frac{1}{\mathcal{N}} \sum_{\mathrm{configs}} \Bigg( \frac{1}{N_{\mathrm{bulk}}} \Big|\sum_{j\in \mathrm{bulk}} s_{j } \Big| \Bigg), \label{eqn:m}
\end{equation}

and the internal energy per spin 
\begin{equation}
    \epsilon=\frac{E}{N_{\mathrm{bulk}}}= \frac{1}{\mathcal{N}}\sum_{\mathrm{configs}} \Big( \frac{1}{N_{\mathrm{bulk}}} \sum_{\langle jk \rangle \in \mathrm{bulk}} s_{j} s_{k}   \Big).\label{eqn:e}
\end{equation}
 
Here, $\mathcal{N}$ denotes the total number of thermalized configurations in the simulation which is $\sim10^7$ sweeps for the Metropolis algorithm and $\sim15000$ sweeps for the cluster algorithm. Each sweep in the metropolis algorithm attempts $N_{0}$ spin flips while each sweep in the cluster algorithm attempts $\sim 15000$ cluster flips. $N_{0}$ is the number of total vertices in the lattice; that is, it includes both what we consider as bulk spins and what we consider as outer layer spins. The susceptibility of the magnetization shown in the Fig.~\ref{bulksus} was computed using the following definition
\begin{align}
     \chi = \frac{\beta}{N_{\text{bulk}}} (\langle M^{2} \rangle - \langle M \rangle^{2}),
\end{align}

and heat capacity per spin was computed with
         \begin{equation}
        C =\frac{\beta^2}{N_{\mathrm{bulk}}} ( \langle E^2 \rangle - \langle E \rangle ^2     ).\label{eqn:chie}
        \end{equation}
        
Plots for the magnetization, internal energy per spin and heat capacity are shown in Fig.~\ref{bulkall}. \\

\section{Correlators of dual spin variables\label{App2}}
In this appendix, we show the boundary correlator of the dual-spin variable $\sigma$ placed on the vertices of the $\{7,3\}$ tessellated disk.  Spin-spin correlator ($\langle \sigma_0 \sigma_r \rangle$) plots at three different temperatures are shown in Fig.~\ref{dual_corr}. Using a similar fitting form to that of Eq.~\eqref{eqn_fit} for the correlators of the disorder variables $\sigma$, we can extract the associated scaling dimension $\widetilde{\Delta}$. We find similar temperature ($\widetilde{T}$) dependence of the scaling exponent $\widetilde{\Delta}$ where the lowest point of a dip correlates to the bulk transition temperature, see Fig.~\ref{dual_Delta}. 


The consistency of our measurements of $\Delta$ and $\widetilde{\Delta}$ can be
checked by noting that the high temperature expansions for the spin-spin correlators on both the dual and direct lattice imply that the correlators take the form
\begin{align}
G_{\rm direct}&\sim e^{R\log\tanh\beta}\sim e^{-2R\widetilde{\beta}}\\
G_{\rm dual}&\sim e^{\widetilde{R}\log\tanh\widetilde{\beta}}\sim e^{-2\widetilde{R}\beta}.
\end{align}
where $R$ is the geodesic distance between spins. 
If we assume the continuum relation $R=\alpha L\log \frac{r}{L}$ with $r$ measured at a finite---effective---boundary, $\alpha$ a constant, and $L$ the AdS radius of the space, these expressions imply the following relations for the scaling dimensions extracted from the boundary correlators
\begin{align}
\Delta&=\alpha L\widetilde{\beta}\quad\widetilde{\beta}\to\infty\\
\widetilde{\Delta}&=\alpha \widetilde{L}\beta\quad \beta\to\infty.
\end{align}
In Fig.~\ref{HighT_fit} we show plots of $\Delta$ v.s. $\widetilde{\beta}$ and $\widetilde{\Delta}$
v.s. $\beta$ including a linear
fit to the large $\widetilde{\beta}$, $\beta$ regions. The ratio of the slopes is
equal to $m_{\rm dual}/m_{\rm direct}=1.97$. If the unit lattice spacing is used for the construction of the Poincar\'e disk, the AdS radius $L$ for a $\{p,q\}$ tessellation
is given by the following formula \cite{mosseri1982bethe}
\begin{equation}
1/L=2 \;{\rm \cosh}^{-1}\left(\frac{\cos{\frac{\pi}{p}}}{\sin{\frac{\pi}{q}}}\right)^2.
\end{equation}
This leads to the prediction $\frac{\widetilde{L}}{L}=1.93$ which agrees remarkably well with the numerical result.


\FloatBarrier
\bibliographystyle{unsrt}

\end{document}